\documentclass[10pt,journal,compsoc]{IEEEtran}
\ifCLASSOPTIONcompsoc
  \usepackage[nocompress]{cite}
\else
  \usepackage{cite}
\fi

\usepackage{cite}
\usepackage{amsmath,amssymb,amsfonts}
\usepackage{algorithmic}
\usepackage{graphicx}
\usepackage{textcomp}
\usepackage{subfigure}
\usepackage{color}
\usepackage[switch]{lineno}

\newtheorem{theorem}{Theorem}[section]
\newtheorem{myDef}{Definition}[section]
\newtheorem{lemma}{Lemma}[section]

\newtheorem{proposition}{Proposition}[section]

\newtheorem{algorithm}{Algorithm}[section]

\ifCLASSINFOpdf
\else
\fi
\begin{document}
%
%
%
%
%

\title{\LARGE \bf
{Resilient Consensus for Multi-Agent Systems \\
 under Adversarial Spreading Processes}}

\author{Yuan~Wang,  \IEEEmembership{Member, IEEE},
        Hideaki~Ishii, \IEEEmembership{Fellow, IEEE},\\
        Fran\c{c}ois~Bonnet,
        and~Xavier~D\'{e}fago, \IEEEmembership{Member, IEEE}
\thanks{%
This work was supported in part
by the JSPS under Grant-in-Aid
for Scientific Research Grant No.~18H01460.
The support provided by the China Scholarship Council is also acknowledged.}
\thanks{Y.~Wang is with
the Division of Decision and Control Systems,
KTH Royal Institute of Technology,
Stockholm, Sweden. H.~Ishii, F.~Bonnet and X.~D\'{e}fago are with the
Department of Computer Science, Tokyo Institute of Technology, Tokyo/Yokohama, Japan. E-mails: wang.y.bb@m.titech.ac.jp,
\{ishii,bonnet,defago\}@c.titech.ac.jp}
}

\markboth{}
{Shell \MakeLowercase{\textit{et al.}}: Bare Demo of IEEEtran.cls for Computer Society Journals}
%



\IEEEtitleabstractindextext{%
\begin{abstract}
This paper addresses novel consensus problems
for multi-agent systems operating in an unreliable
environment where adversaries are spreading.
The dynamics of the adversarial spreading processes follows the
susceptible-infected-recovered (SIR) model,
where the infection
induces faulty behaviors in the agents
and affects their state values. Such a problem setting serves as a model of
opinion dynamics in social networks where consensus
is to be formed at the time of pandemic and infected
individuals may deviate from their true opinions.
To ensure resilient consensus among the noninfectious
agents, the difficulty is that the number of
infectious agents changes over time.
We assume that a local policy maker announces
the local level of infection in real-time, which can
be adopted by the agent for its preventative measures.
It is demonstrated that this problem can be formulated
as resilient consensus in the presence of
the socalled mobile malicious models,
where the mean subsequence reduced (MSR) algorithms
are known to be effective.
We characterize sufficient conditions on the network
structures for different policies regarding the
announced infection levels and the strength of the
epidemic.
Numerical simulations are carried out for
random graphs to verify the effectiveness
of our approach.
\end{abstract}

}

\maketitle

\IEEEdisplaynontitleabstractindextext

%
\IEEEpeerreviewmaketitle

\vspace{-2 ex}

\section{Introduction}

Recently, the pandemic of COVID-19 has highlighted the necessity
and effectiveness of unknown disease peak control.
Since it may take up to several years to develop and supply
vaccines sufficiently broadly \cite{Graham2018},
a large ratio of population may become infected simultaneously,
which would cause huge pressure to hospitals and medical sectors.
Initial measures such as keeping social distance and avoiding
unnecessary gatherings are introduced to slow down the
disease spreading. By temporarily reducing contacts
with others at proper periods, peaks can be
reduced in frequency and the number of infected patients.
Recent studies related to theoretical studies on
peak control can be found in, e.g, \cite{Morris2020,Lauro2020}.

In this paper, we are interested in studying the impact of
adversarial spreading processes in the context of multi-agent systems.
In such systems, agents locally interact with each other to carry
out certain global tasks, but infected agents may deviate from
their normal behaviors, which may even be harmful to the system.
Hence, the healthy agents must pay attention to the level of epidemics
in the environment and accordingly regulate their interactions with others.

As the global task of the agents, we limit our study to the
problem of consensus, which is also important in applications such as
those in opinion dynamics from social networks \cite{DeGroot1974,Hegselmann2002,Blondel2009,Su2017}.
Under conventional consensus algorithms, it is well known that
agents can become influential by simply being stubborn, keeping
their state values unchanged. This means that infected agents may falsely lead the state values of others. In the worst case,
they may split the agents into several clusters in their states and prevent
them from consensus forming.

In particular, we formulate a novel problem
where the agents may fall in infectious statuses
depending on the susceptible-infected-recovered (SIR)
epidemic dynamics in the environment. Infected agents will
take unexpected behaviors in their values and agents free
from the infections should avoid using such state values
when updating their own ones. Here, we employ
resilient versions of consensus algorithms, which are designed
to be used in the presence of adversaries, proposed in the works
such as \cite{LeBlanc2013,Vaidya2012,wang:ifac2020}.
While we follow the line of research on fault tolerant
consensus algorithms (e.g., \cite{LeBlanc2013}), the main difference is that
the number of faulty agents is time varying.
\textcolor{black}{This information
may be partially known from a local policy maker
who attempts to estimate the epidemic condition and makes
announcements for the local area about it in real time. For example, medical experts
may announce the current statistics of the number of patients through the media;
we however point out that a policy maker will not assist
the consensus forming that locally takes place
among agents.}
We believe that the studied problem is helpful to enhance the resilience for
consensus-based applications under adversarial spreading
processes such as opinion dynamics at the time of pandemic.
To the best of our knowledge, this paper is the first dealing
with such a problem.

It is emphasized that to deal with this problem,
the notion of mobile malicious agents studied
in \cite{wang:ifac2020}
becomes critical. {Most existing works on resilient
consensus focus on adversaries that remain fixed at certain agents.} Inspired by epidemic peak control
\cite{Morris2020,Lauro2020},
we extend the mobile malicious model to the case with
pandemics.
Here, the malicious agents no longer move, but may infect other
regular agents. For the infected agents, their behaviors as well
as their values can be changed. Meanwhile, the infected agents will
also recover from the infection at a certain rate.
Once cured, they should follow the designed update
rules even if their own values may be corrupted.

\emph{Related works: }
Epidemic spreading models have been long studied among many
different fields including mathematical biology
\cite{Kermack1927}, computer science \cite{Ostrovsky1991,Serazzi2003,Yoshiaki2014},
social science \cite{Hegselmann2002,Kempe2003,Blondel2009,Su2017,She2021} and so on.
Various epidemic models have been studied,
but the most common ones include the susceptible-infected-susceptible
(SIS) model and the SIR model.
In the traditional SIS model, the agents are divided into
two groups taking
the susceptible and infected states; the ratios of the two groups may
exhibit dynamic behaviors over time.
More recently, the SIS model has been incorporated with networks representing
interactions of agents, where the pandemic process evolves over
time-varying networks \cite{Pare2017, Pare2019, Vrabac2020}.
Similar trends can be found in the SIR model, where the agents may
be in susceptible, infected, or recovered state.
Once agents are recovered from their infected states,
they will not be infected again.
Recent works on SIR-type models focus on
improving the model based on real data \cite{Chen2020},
considering time delay issues \cite{Briat2008},
and applying the SIR model in other problems such as
information source detection \cite{Chen2016} and
information epidemics in social networks \cite{Liu2020}.
In \cite{Nowzari2016}, the development, analysis and
control problems for epidemic models are reviewed from
the viewpoint of systems control.  {More recently, several works attempt to combine opinion dynamics
models with epidemic models in, e.g., \cite{She2021,Xuan2020},
where the two models are assumed to have the same network structure and the coupling in their dynamics is highlighted.}

In contrast, our work focuses on consensus forming and
follows the line of fault-tolerant algorithms for
multi-agent systems. We assume that the infected agents follow the so-called \emph{malicious adversary model}. The infected agents may lose their original state values and broadcast their corrupted values. The goal for other non-infected agents is to reach consensus at a safe value, within the range of the original values of the non-infectious agents. Moreover, our results are motivated by the class of Mean Subsequence Reduced (MSR) algorithms, which has been studied in \cite{Vaidya2012,LeBlanc2013,wang:ifac2020}. In such algorithms, the non-infected agents ignore suspicious values from their neighbors. Such algorithms do not need to detect the adversarial neighbors, but require a certain level of connectivity in the network. {
In the area of opinion dynamics, MSR-like algorithms can be
found such as the Hegselmann-Krause (HK) model \cite{Hegselmann2002,Blondel2009,Su2017}.
There, each agent removes all values that are sufficiently different from its own opinion at eachtime step before taking the average of the rest in their updates.
Different from the MSR algorithm, the number of removed values is not restricted. }

{
\emph{Other applications}:
There are other interpretations of the proposed two-layer
multi-agent consensus model under adversarial spreading processes
studied in this paper. One is to consider
the epidemic and the behaviors of the malicious agents
as outcomes of rumors spreading in the community. In fact rumors
can be viewed as an ``infection of the mind.''
Studies on transmission of rumors over the Internet
and the behaviors of social networks can be found in, e.g.,
\cite{Nekovee2007,Zhao2013, Jin2013}. In this setting, one may obtain the information
about the level of rumors spreading from SNS and other sources.
}

{
On the other hand, in the context of computer networks and wireless sensor networks,
propagation of viruses may follow epidemic models as
discussed in \cite{Serazzi2003,Yoshiaki2014,De2009}. Indeed, distributed algorithms
for consensus have a range of applications such as load balancing in multiprocessor networks \cite{Amelina2015}, averaging \cite{Chamie2016},
clock synchronization in wireless sensor networks \cite{Kikuya2018},
and rendezvous in robotic networks \cite{Park2016}.
Cautions for viruses may be transmitted by security software
companies, to which individual agents can adopt.
}


\emph{Contributions:}
The contribution of this work is threefold:
First, we extend the mobile adversary model
of one-to-one mobile behavior to an epidemic case.
In our previous work \cite{wang:ifac2020},
a malicious agent moves to another agent and leaves a
corrupted value.
The agent that just recovered is considered as a cured agent.
Note however that such an agent may have a corrupted value
and hence should be treated as a malicious one for another round
so as to apply a protocol to adjust its value.
Second, we introduce an intervention measure to the epidemic model
to enhance the resiliency of the consensus protocols
employing the modified MSR algorithms.
In conventional epidemic works such as \cite{Morris2020},
the intervention takes the form of the ratio at which
the general public should avoid contacts with others.
In this work, we relate such intervention ratio
with the parameters employed in MSR algorithms
for pruning the neighbors' values;
we formally analyze their role in resilient consensus.
Third, based on the modified MSR algorithms
used for mobile malicious models, we propose
two protocols with static and adaptive policies
for the intervention ratio globally announced
and hence the the pruning parameters locally
at the agents.
For both protocols, we characterize the graph conditions
and tolerable pandemic level under which resilient consensus
can be guaranteed.
Compared with the conventional mobile malicious model,
the epidemic malicious model is more powerful since it
has dynamic malicious agents. The pruning number
is no longer given, but has to be chosen properly
according to the infection level at the time.
The existence of pruning numbers that can guarantee
resilient consensus in a given graph is also discussed.

\emph{Outline: }
This paper is organized as follows.
In Section~\ref{Section 2},
preliminaries and the general problem
setting are introduced.
In Section~\ref{sec:agents},
two protocols for resilient consensus
are proposed based on static and realtime information
regarding the infection.
Our results for complete and noncomplete graphs
are presented
in Sections~\ref{Section 4} and~\ref{Section 5}, respectively.
We demonstrate the efficacy of the
algorithms by numerical examples in
Section~\ref{Section 6}.
Section~\ref{Section 7} gives concluding remarks.
A preliminary version of this paper will appear as a conference paper \cite{wang:ecc2021}.
The current paper contains all proofs for theoretical results and extensive simulations are carried out as well.


\section{Problem Formulation}
\label{Section 2}

\subsection{Preliminaries on Graphs}


Denote by $\mathcal{G}=(\mathcal{V},\mathcal{E})$ the graph
consisting of $n$ nodes, where $\mathcal{V}=\{1,2,\ldots,n\}$ is
the set of nodes and
$\mathcal{E} \subset \mathcal{V} \times \mathcal{V}$ is the set of edges.
The edge $(j,i) \in \mathcal{E}$ indicates that node~$j$ can send
a message to node~$i$ and is called an incoming edge of node~$i$.
Directed graphs are considered, in which $(j,i) \in \mathcal{E}$
does not necessarily imply $(i,j) \in \mathcal{E}$.
Let $\mathcal{N}_{i}=\{j:(j,i)\in \mathcal{E}\}$
be the set of in-neighbors of node $i$.
The in-degree $d_i$ of node~$i$ indicates the number
of its in-neighbors, i.e., $d_i=|\mathcal{N}_{i}|$.

The path from node~$i_1$ to node~$i_p$ is denoted as
the sequence $(i_1, i_2, \ldots, i_p)$, where
$(i_j,i_{j+1}) \in \mathcal{E}$ for $j=1, \ldots, p-1$.
The graph $\mathcal{G}$ is said to contain a spanning tree
if from some node, there are paths to all
other nodes in the graph.
\textcolor{black}{The graph $\mathcal{G}$ is partitioned into $m$ disjoint subgraphs and we denote them by $\mathcal{G}_s=(\mathcal{V}_s,\mathcal{E}_s), s \in \{ 1, \ldots, m \}$, where $m \le n$, $\mathcal{V}_s \cap \mathcal{V}_r = \varnothing$ for all $s,r \in \{ 1, \ldots, m \}$, $\mathcal{V}_1 \cup \cdots \cup \mathcal{V}_m = \mathcal{V}$ and $\mathcal{E}_s \subset \mathcal{V}_s \times \mathcal{V}_s$.}

\subsection{Overall System Model}

In this subsection, we provide an overview of the problem setting of
resilient consensus for multi-agent systems in an environment
where adversaries are spreading.

\begin{figure}[t]
\centering
  \includegraphics[width=1\linewidth]{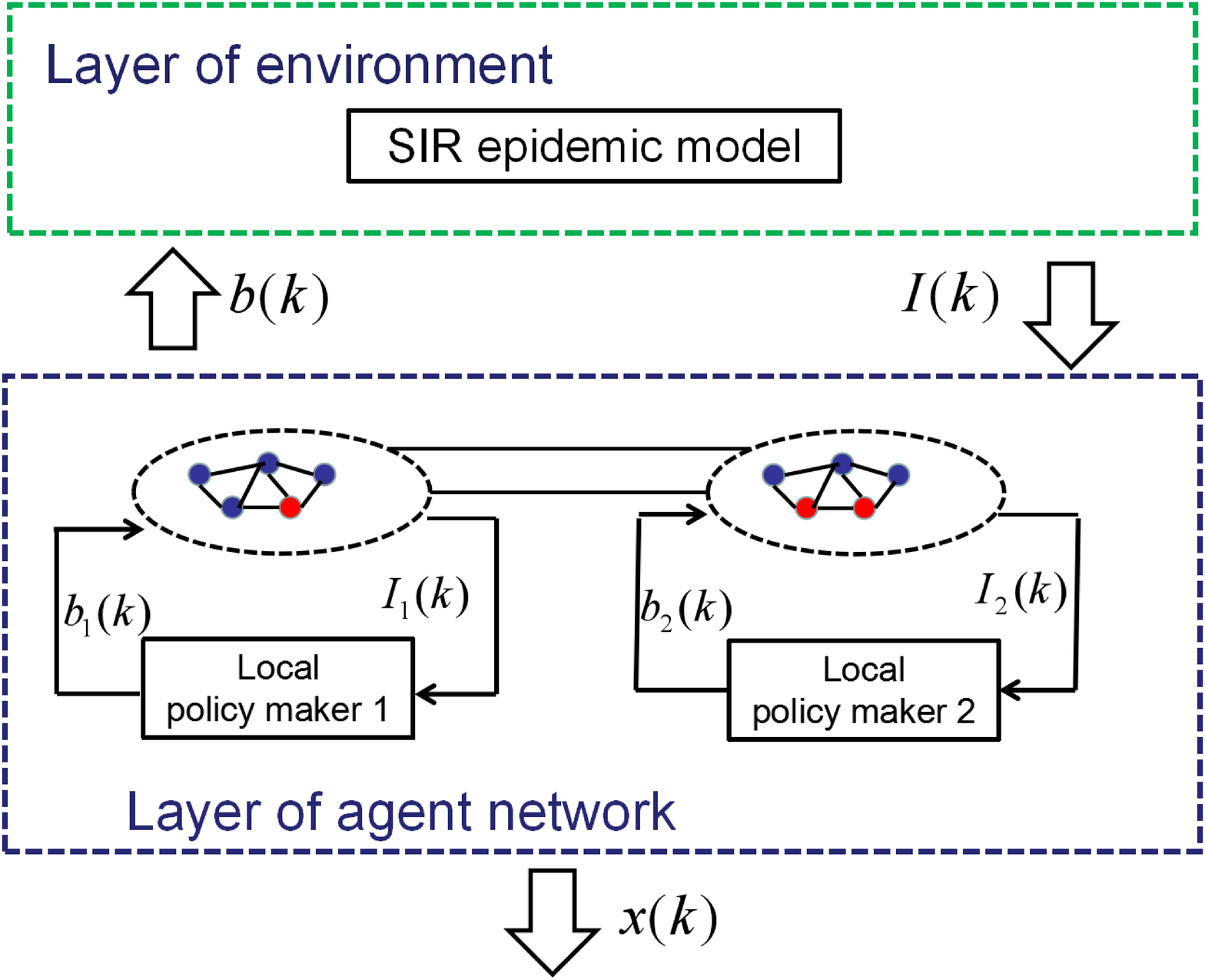}
  \vspace*{-6mm}
  \caption{Feedback architecture for the layer of environment and layer of agent network with $m=2$}
  \vspace*{-4mm}
\label{fig_sys}
\end{figure}

The overall system considered here consists of two layers,
representing different models related to (i)~the environment
and (ii)~multi-agent systems. \textcolor{black}{This is shown in Fig.~\ref{fig_sys}, where the two layers interact with each other and there is
an overall feedback structure.
The layer of the environment is determined by the SIR epidemic model.
In the agent network layer, there are multiple local policy makers placed in the subgroups. They can estimate the ratio of the infection in the subgroup and then to announce how every agent in the subgroup should reduce their interactions with others. The announcement is made by adjusting the socalled reduction parameter.}

\subsection{First Layer with the SIR Spreading Process}

In the first layer, to describe the adversarial spreading,
\textcolor{black}{we consider a variant of the standard SIR spreading model with policy makers who attempt to regulate spreading of the adversaries
by making announcements so that the possible malicious contacts within the environment decrease
\cite{Morris2020}.}

In the SIR model, the average fractions of the agents susceptible,
infectious, and recovered at continuous time are denoted,
respectively, by $S(t)$, $I(t)$, and $R(t)$, where $S(t)+I(t)+R(t)=1$ at all $t$.
The adversaries spread at the transmitting
rate $\beta > 0$ while the infection recovers at
the recovering rate $\gamma > 0$.
It is common to denote the basic
reproduction number by $R_0 = \beta/\gamma$.
It represents the reproduction ability, which may correspond
to the strength of the general adversarial spreading behavior.

\textcolor{black}{The role of the policy makers is to regulate the transmitting rates by forcing all the local members to reduce their contacts with
others. The overall transmission reduction rate is represented by the global
parameter $b(t)\in[0,1]$. It can be regarded as the overall feedback control input from the agent network layer. This results in a smaller transmitting rate at $b(t)\beta$. In the conventional SIR model where no regulation is made, this parameter remains at $b(t)\equiv 1$.}


With the transmission reduction based on $b(t)$ at time $t$, the pandemic process is described by the SIR model as
\begin{equation}\label{eq-new01}
\begin{split}
    \dot{S}(t) &= - b(t)\beta S(t)I(t),  \\
    \dot{I}(t) &= b(t)\beta S(t)I(t) - \gamma I(t),  \\
    \dot{R}(t) &= \gamma I(t).
\end{split}
\end{equation}
In this paper, we deal with the system
in the discrete-time domain by discretizing the SIR model above.
Let $\Delta T$ be the sampling period.
Denote the variables at time $k\Delta T$
by $S(k)$, $I(k)$, and $R(k)$ and so on. {For simplicity,
we take the sampling period for the epidemics
to be the same as that of the agents. }
Based on the Euler method, when $\Delta T$ is
sufficiently small, the continuous-time
dynamics in \eqref{eq-new01} can be approximately
described by the following discrete-time dynamics:
\begin{equation}\label{eq-01}
\begin{split}
    S(k+1) &= S(k) - b(k)\beta S(k)I(k)\Delta T,  \\
    I(k+1) &= I(k) + b(k)\beta S(k)I(k)\Delta T - \gamma I(k)\Delta T,  \\
    R(k+1) &= R(k) + \gamma I(k)\Delta T.
\end{split}
\end{equation}
As mentioned above, $b(k)$ is the
key parameter to control the pandemic
peak. It is determined by the policy
maker who performs the analysis on $I(k)$.

\begin{figure}[t]
\centering
  \includegraphics[width=1.08\linewidth]{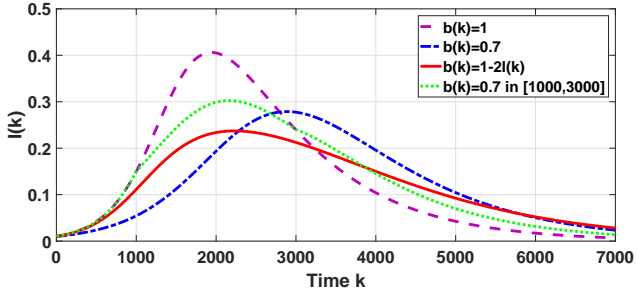}
  \vspace*{-6mm}
  \caption{Infectious ratio $I(k)$ under
            different control policies}
  \vspace*{-4mm}
\label{fig00}
\end{figure}

Different policies can be considered for the choice of the
transmission reduction. We illustrate this point by
a numerical example with the basic parameters taken as
$\beta=0.4$, $\gamma=0.1$, and $\Delta T =0.01$.
In Fig.~\ref{fig00}, the time responses of the
infectious ratio $I(k)$ are shown under four policies:
(i)~The base case with no reduction (i.e., $b(k)\equiv1$) in dashed line,
(ii)~fixed reduction at $b(k)\equiv 0.7$ in dash-dot line,
(iii)~adaptive reduction $b(k)=1-2I(k)$ with the knowledge
of the infectious rate $I(k)$ in solid line, and
(iv)~reduction at limited time period based on
$b(k)=0.7$ for $k\in[1000,3000]$ and 1 otherwise
in dotted line.
The initial states are taken as $S(0)=0.99$, $I(0)=0.01$,
and $R(0)= 0$.

Notice that in the base case (i), the transmitting rate $\beta$ is
high, resulting in the peak of $I(k)$
greater than $0.4$. In the other cases, the peaks are
about $0.3$. The simple policy (ii) with $b(k)\equiv 0.7$ is the most
demanding in terms of transmission reduction over time;
the peak is smaller than $0.3$, but occurs late,
delaying the recovery. The adaptive policy (iii) is also demanding,
but the reduction increases slowly in
response to $I(k)$; the peak is less than 0.25
and occurs early. The time-limited method (iv) is also effective to keep the peak to appear early in time.

\subsection{Second Layer of the Multi-agent Network}

The second layer is at the lower level, containing a number of agents
represented by a network.
\textcolor{black}{As described above, the adversarial dynamics is represented
by the SIR epidemic model. This determines the ratios of agents
that are susceptible, infectious, and recovered. Now, we partition the graph into $m$ subgraphs $\mathcal{G}_1,\ldots,\mathcal{G}_m$.
The agents in each subgroup are subject to infections though
the infection ratios may be different in the subgroups.
Moreover, each subgroup is
equipped with a local policy maker who makes the local
announcement for real-time transmission reduction in the
subgroup. Since the agents in the subgroups follow the
announcements, the overall infection dynamics in the
environment is regulated as we shall see more later. }

\textcolor{black}{The agent set $\mathcal{V}$ is
partitioned into three disjoint sets: $\mathcal{S}(k)$ of susceptible agents,
$\mathcal{I}(k)$ of infectious agents, and $\mathcal{R}(k)$
of recovered agents, where
$\mathcal{S}(k)\cup\mathcal{I}(k)\cup\mathcal{R}(k)=\mathcal{V}$.
Denote the local ratio of the infection $I_s(k)$ by
\begin{equation}\label{eq-Is}
I_s(k) = \frac{ \big| \big( \bigcup_{i\in \mathcal{V}_s} \mathcal{N}_i \big) \cap \mathcal{I}(k)\big| }{ \big| \bigcup_{i\in \mathcal{V}_s} \mathcal{N}_i \big| }, s \in \{ 1, \ldots, m \}.
\end{equation}
We consider the heterogenous case where $I_s(k)$ is different from subgroups. The global ratio of infection $I(k)$ is
\[
I(k) = \frac{|\mathcal{I}(k)|}{|\mathcal{V}|}.
\]
}
\textcolor{black}{
Then for all $s \in \{ 1, \ldots, m \}$, the local infectious ratio dynamic can be can be described by
\begin{equation}\label{eq-01l}
    I_s(k) = I(k) + w_s(k),
\end{equation}
where $|w_s(k)| \le \overline w \in [0,1]$ is the real time heterogeneity for subgraph $\mathcal{G}_s$. Clearly, if $\overline w =0$, then it is a homogenous spreading process for all subgroups.
}

\textcolor{black}{In the agent network layer, there is a local policy maker for each subgroup $s$ in charge to
estimate the local ratio of the infection $I_s(k)$ and
then to announce how the agents in subgraph $\mathcal{G}_s$ should
reduce their interactions with others.
The announcement is made by adjusting the socalled
\textit{local reduction parameter} $b_s(k)$. The global reduction parameter $b(k)$ is the key parameter that provides a feedback control mechanism to suppress the infection peak in the SIR spreading process. It is represented by the interval containing all the local reduction parameters $b_s(k)$ and thus given by $b(k) \in [\min_{s \in \{ 1, \ldots, m \}}{b_s(k)},\max_{s \in \{ 1, \ldots, m \}}{b_s(k)}]$.}


Furthermore, in the second layer, the regular agents
in the susceptible and recovered states employ a resilient
variant of consensus protocols called the MSR algorithms.
{In the application example of opinion dynamics under pandemic, the states of the agents represent the opinions of the individuals on certain issues (social, political, and so on).}
Each agent will follow the announced level of reduction parameter from its corresponding local policy maker.
Specifically, they reduce their contacts by ignoring some of
the values received from neighbors.
Notice that due to the SIR epidemic model,
the number of neighbors who may be infectious is time varying.
In particular, the recovered agents have to be careful not
to restart their consensus protocols using the values left from
the infected periods.
Under such circumstances,
 conventional resilient consensus algorithms
are not capable to eliminate the adversarial effects. Instead,
one must resort to the notion of mobile adversarial models,
where the identities of the malicious agents switch,
and also to algorithms robust to such models
(e.g.,\cite{Buhrman1995, Garay1994,wang:ifac2020}).

The objective of the multi-agent system in the lower layer
is to reach consensus on their state values, and this
must be achieved without being influenced by the infected
agents. The susceptible agents may not become infected by
solely interacting with infectious agents. However,
the states of the infectious agents may be
affected by the adversaries.
Thus, the susceptible
and recovered agents must take preventative measures
at the time of updating their own state values.

In our problem setting, three issues are present, creating
difficulties for the decision making of the agents to reach
consensus. First, the infection may spread quickly
depending on the parameters that determine the strength
of the adversarial spreading, and large peaks may appear.
Second, {the ratio of infectious agents} is unknown
in general.
In the second layer, this means that
the identities of the infectious agents
are unknown to the susceptible and recovered ones.
As a third hurdle, we impose that the agents must
continue with their interactions during the high peaks.
Note that if sufficiently many agents become infected over time,
the original values of their states may become lost from the system,
which will make it difficult for the agents to arrive at a
safe, reasonable value for consensus.
We mention this point since
in the SIR epidemic model, the infections are bound to cease
in the long run. Hence, the most critical incident
that must be avoided is to lose safe values
from the entire multi-agent system.

\section{Resilient Consensus in the Layer of the Multi-agent System}
\label{sec:agents}

In this section, we explain the resilient consensus in the lower layer of the overall system, where
the multi-agent system is placed.

\vspace{-2 ex}
\subsection{Resilient Consensus}
The network of agents is
represented by the directed graph $\mathcal{G}$.
In our setting, the status of the agents is determined
by the condition of the adversaries in the environment as follows.
At each time $k$, in accordance with the fractions of $S(k)$,
$I(k)$, and $R(k)$ in the SIR epidemic model.

Each agent $i\in\mathcal{V}$ has a state value denoted
by $x_i(k)$ at time $k$.
At nominal times when no outbreak of the adversaries is present,
all agents would follow the protocol below for seeking consensus
in their values:
\begin{equation}\label{eqn:consensus}
 x_i(k + 1)
  = \sum\limits_{j \in \{i\}\cup\mathcal{N}_{i}}
              a_{ij}(k) x_j(k),
\end{equation}
where the weights $a_{ij}(k)\in[0,1)$ satisfy $a_{ii}(k)+\sum_{j \in \mathcal{N}_{i}} a_{ij}(k)=1$.
It is well known that the agents will arrive at consensus,
i.e., $\bigl|x_i(k)-x_j(k)\bigr|\rightarrow 0$ as $k\rightarrow \infty$
for $i,j\in\mathcal{V}$
if the network $\mathcal{G}$ contains a directed spanning tree.

We call the agents to be \textit{regular} if they are in the susceptible status
$\mathcal{S}(k)$ and the recovered status $\mathcal{R}(k)$. These agents are capable to execute
the given algorithm and maintain their values accordingly.
On the other hand, the infected agents may have corrupted values.
In particular, for each agent
in the infectious status $\mathcal{I}(k)$, its value is updated as
\begin{equation}\label{eqn:infected_agent}
   x_i(k+1) = u_i(k),~~i\in\mathcal{I}(k),
\end{equation}
where the input $u_i(k)$ can be set arbitrarily due to the
adversaries.

Now, we introduce the notion of resilient consensus
for the multi-agent system under the epidemic model.

\begin{myDef}\label{def5}
 (\emph{Resilient consensus}) If for any possible sets
and behaviors of the infectious agents and any state values
of the regular agents, the following two conditions are
satisfied, then we say that the regular agents reach
resilient consensus:
\begin{enumerate}
 \item Safety condition:
   There exists a bounded interval
   $\mathcal{B} \subset \mathbb{R}$ determined by the
   initial states of the regular (susceptible and recovered) agents
   such that $x_i(k)\in \mathcal{B}$ for all
   $ i \in \mathcal{S}(k) \cup \mathcal{R}(k)$, $k\in \mathbb{Z}_{+}$.
 \item Consensus condition:
   The regular agents eventually take the same value as
  $\max_{i,j\in\mathcal{S}(k) \cup \mathcal{R}(k)}
      \bigl| x_i(k)-x_j(k)\bigr|\rightarrow 0$
  as $k\rightarrow\infty$.
\end{enumerate}
\end{myDef}

{
Under adversarial spreading,
it is possible that resilient consensus is achieved
before adversaries die out, but as long as
spreading adversaries remain in the system,
the values of the agents in the normal status, i.e. those in the set $\mathcal{S}(k) \cup \mathcal{R}(k)$, may change over time.
This is because agents that just recovered can have corrupted values, and when they rejoin the system,
it may take time for the normal agents to reach consensus again.
However, in Definition~\ref{def5} above, resilient consensus means that
such transient behaviours will eventually stop.
The definition is consistent with that for the case
of adversaries at fixed agents as in \cite{LeBlanc2013,Dibaji2018}, where once the regular agents achieve resilient consensus,
the agents will remain in that status.
}

%

In our approach, the regular agents
follow the socalled MSR algorithm to protect their values
from being corrupted by using those received from infected agents.
Their states are updated as in \eqref{eqn:consensus},
but with a restricted number of neighbor values
using the pruning number $F_i(k)\in[0,d_i/2)$
for $i\in\mathcal{V}$.
The algorithm is outlined below.
In particular, we present a modified version of the MSR
algorithm from \cite{wang:ifac2020}.

\begin{algorithm}\label{protocol}
At each round $k$,
each regular agent~$i\in \mathcal{S}(k) \cup \mathcal{R}(k)$
executes the following three steps:
\begin{enumerate}
  \item Agent~$i$ sorts the values ${x}_j(k)$, $j\in \mathcal{N}_i$,
        received from its neighbors
        and its own value ${x}_i(k)$ in descending order.
  \item After sorting, agent~$i$ deletes the $F_i(k)$ largest
        and the $F_i(k)$ smallest values.
        The deleted data will not be used in the update.
        The set of indices of agents whose values remained
        is written as $\mathcal{M}_i^+(k) \subset \{i\}\cup\mathcal{N}_i$.
  \item Finally, agent~$i$ updates its value by
      \begin{equation}\label{eq-02}
        {x_i}(k + 1)
         = \sum\limits_{j \in \mathcal{M}_{i}^+(k) }
               a_{ij}(k) x_j(k).
      \end{equation}
\end{enumerate}
\end{algorithm}

\textcolor{black}{We would like to introduce the following locally homogenous assumption as the first step of this study.
The local infection ratio $I_s(k)$ represents the infected ratio around each agent in subgroup $s$:
\begin{equation}\label{eqn:Ho}
  \left|
      \mathcal{N}_i \cap \mathcal{I}(k)
  \right|
     \le d_i I_s(k), i \in \mathcal{V}_s.
\end{equation}
}
\textcolor{black}{Note that this assumption indicates that the infection ratio around each agent in the subgraph $\mathcal{G}_s$ is homogenous. However, as a special case, if $|\mathcal{V}_s| = 1, \forall s \in \{ 1, \ldots, m \}$ (i.e., each agent is equipped with a policy maker), then based on \eqref{eq-Is}, we always have \eqref{eqn:Ho}. This locally homogenous assumption is clearly an assumption that simplifies the problem, and it is left for future studies to use more sophisticated models \textcolor{black}{such as \cite{Pare2017,Trajanovski2015}}.}

{To design a successful MSR algorithm to prevent the corrupted values
from affecting the agents, it is critical that
the regular agents have sufficient information regarding the ratio of local infections. In particular, the designer must guarantee that all the local pruning number
$F_i(k)$ is greater than
the number of infected neighbors and smaller than half of the local degree at all $k$ as
\begin{equation}\label{eqn:Fi0}
  \left|
       \mathcal{N}_i \cap\mathcal{I}(k)
  \right| \le F_i(k) < \frac{d_i}{2}, ~~i\in\mathcal{S}(k)\cup\mathcal{R}(k).
\end{equation}
Otherwise it is impossible to remove all the corrupted values. }
Note however that the exact information of
$\mathcal{I}(k)$ is not available to anyone in the system.
\textcolor{black}{In our approach, we connect
the pruning number $F_i(k)$ used in the MSR algorithm and
the local transmission reduction parameter $b_s(k)$, where $i \in \mathcal{V}_s$.
This point is discussed in the next subsection.}

\vspace{-2 ex}

\subsection{Infectious Rate and Transmission Reduction}


\textcolor{black}{The local policy maker has an estimate of the ratio $I_s(k)$ of local infectious agents
and decide the local transmission reduction parameter $b_s(k)$.
Note that $1-b_s(k)$ represents the ratio of contacts that
node $i$ should cut down on. The nominal case without infected agents means $b_s(k)=0$ and all neighbor values can be used. On the other hand, there are at most ${d}_i I_s(k)$ corrupted values based on \eqref{eqn:Ho}.
Under the modified MSR algorithm, node $i \in \mathcal{V}_s$ should remove $2{d}_i I_s(k)$ suspicious values, which means $b_s(k) = 1 - 2I_s(k)$. Considering the bound for subgroup heterogeneity $\overline w$ in \eqref{eq-01l} in addition, we assume that the local policy maker chooses the local transmission reduction parameter $b_s(k)$ by
\begin{equation} \label{eqn:bk}
  b_s(k)\in[0, 1 - 2I_s(k)-2 \overline w].
\end{equation}
}
Hence, upon receiving the announced value of $b_s(k)$,
to follow the policy maker,
the regular agents need to choose their pruning numbers $F_i(k)$
such that $2F_i(k)/d_i \ge 1-b_s(k)$.
Moreover, $F_i(k) < d_i/2$ is necessary
for agents using the MSR algorithm to have
at least one neighbor at each update after
the removal of
extreme values. Thus, $F_i(k)$ must be chosen as
\begin{equation} \label{eqn:Fi2}
   F_i(k) \in \Bigl[
                \left(1-b_s(k)\right)\frac{d_i}{2},\frac{d_i}{2}
              \Bigr),
   ~~i\in \big(\mathcal{S}(k) \cup \mathcal{R}(k)\big) \cap \mathcal{V}_s.
\end{equation}
\textcolor{black}{In fact, from \eqref{eqn:Ho}, \eqref{eqn:bk} and \eqref{eqn:Fi2},
we can obtain the bound \eqref{eqn:Fi0}.}

\textcolor{black}{
Moreover, by local transmission reduction parameter $b_s(k)$ in \eqref{eqn:bk}, based on~\eqref{eq-01l} and $b(k) \in [\min{b_s(k)},\max{b_s(k)}]$, we have
\begin{equation} \label{eqn:bk-general}
  b(k)\in [0, 1 - 2I(k)].
\end{equation}
}
\vspace{-2ex}

It is important that  each local policy maker of node $i$ is sufficiently
knowledgeable that he always selects the transmission
reduction parameter $b_i(k)$ satisfying \eqref{eqn:bk}. As long as
this condition is met, the frequency of the updates in $b_i(k)$
need not be the same as that of the agent states.

\vspace{-2ex}
\subsection{Model for Infectious Agents}

Here, we characterise the model for the infectious agents
in terms of their behaviors when they are infected and then
recovered.
The infectious agents are considered to be adversarial
in this work. In particular, we follow
the malicious model of \cite{LeBlanc2013}, where
the classification is based on their number,
locations, and behaviors.

\begin{myDef}\label{def3}
(\emph{Malicious agents}) An infected agent~$i\in\mathcal{I}(k)$
is said to be malicious if it can arbitrarily modify its local
variables as in \eqref{eqn:infected_agent}
and send the same value to all of
its neighbors each time a transmission is made.
\end{myDef}

{The motivation for considering malicious agents as defined above comes,
for example, from the applications of
social networks and computer networks where agents
communicate by broadcasting their data. Especially the
infected agents might loudly announce their extreme opinions/data to their neighbors.}.

It is important to notice
that under our pandemic model, the identities
of the infected, malicious agents change over time.
This is in contrast to the conventional models
in, e.g., \cite{LeBlanc2013},
where the malicious agents remain the same.
To this end, we must incorporate the more general model known as
the mobile malicious agents studied in, e.g.,
\cite{Buhrman1995, Garay1994}.

\begin{figure}[t]
\centering
\includegraphics[width=1\linewidth]{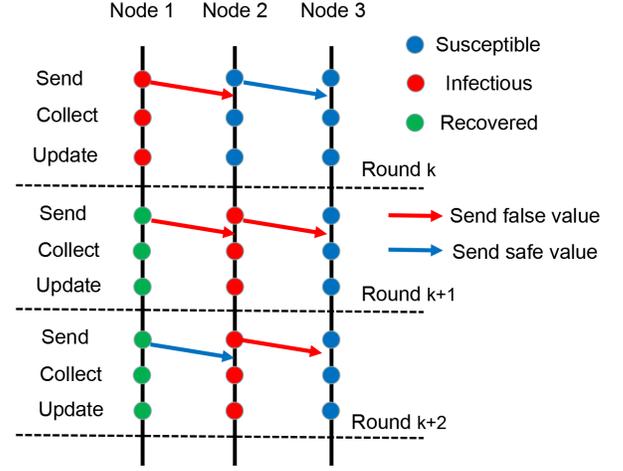}
  \vspace*{-2mm}
\caption{Epidemic malicious model}
  \vspace*{-4mm}
\label{figadd}
\end{figure}

Under such mobile malicious models, the infectious agents have
two properties different
from the conventional static model (see Fig.~\ref{figadd}).
First, a malicious agent may infect regular agents so that their
statuses change.
While infected, agent~$i\in\mathcal{I}(k)$ broadcasts its corrupted
state $x_i(k)$ (controlled as in \eqref{eqn:infected_agent})
to its neighbors, but then becomes
recovered at the recovering rate in the
SIR epidemic model.
Second, once recovered, agent~$i\in\mathcal{R}(k)$ collects
and updates its own as a regular agent. However, in the
first round after the recovery,
the agent may still possess a corrupted value
left from
its infected period. Hence, such an agent should be considered
still infected and will be said to be in the cured status.
Moreover, the agent should refrain from using
its own value in the cured status.
These aspects will be taken into account in the proposed
algorithm.

\vspace{-2ex}

\subsection{Control Policies under the Epidemic Model}

In this paper, we address the resilient consensus problem
where the agents' statuses change based on the epidemic model.
Regarding the local transmission reduction parameter $b_i(k)$,
two policies are studied: Static and dynamic.
The analyses for the two policies will follow in Section~\ref{Section 4}
for the complete graph case and
Section~\ref{Section 5} for the noncomplete graph case.


\section{Resiliency Results for Complete Graphs}
\label{Section 4}

We establish conditions under which
the agents equipped with Algorithm~\ref{protocol}
can reach resilient consensus
in the epidemic malicious model.
In this section, we first present the results
for networks in a complete graph, \textcolor{black}{where all nodes interact with each other. Hence, from \eqref{eq-Is}, there is no subgroup heterogeneity and $\forall s \in \{1,\ldots,m \}, b_s(k)= b(k)$, $I_s(k) = I(k)$ in this section.}

The static policy is a special case, where
the local transmission reduction parameter is fixed as
$b(k)\equiv b_{0} \in[0,1]$ for the entire time horizon.
This constraint may limit the feasible strategies
for the overall system since the condition
in \eqref{eqn:bk} reduces to
$b_{0}\leq 1 - 2I(k)$ for all $k$.
{As a consequence, from~\eqref{eqn:Fi2}, each regular agent~$i$ can use
a constant for the number $F_i(k)\equiv F_{i0} \ge (1-b_{0}){d_i}/{2}$
of neighbors to be removed as well.}

In fact, for this case, an analytic bound for the pandemic
peak can be obtained.
Suppose that the local transmission reduction parameter $b_{0}$
is large enough that
\begin{equation}\label{bR0}
  b_{0} R_0 > 1.
\end{equation}
Note that this relation requires $R_0>1$.
From the work \cite{Kermack1927}, it is known
that under these conditions,
the maximum of $I(t)$ in \eqref{eq-new01} can be obtained as
\begin{equation*}
  \max_{t>0} I(t)
    = I(0) + S(0) - \frac{1}{b_0 R_0}\ln S(0)
        - \frac{1}{b_0 R_0}
        + \frac{1}{b_0 R_0}\ln \frac{1}{b_0 R_0}.
\end{equation*}
Since $I(k)$ is the sampled value from \eqref{eq-new01} and the sampling period $\Delta T$ is small enough,
approximate of the maximum $I(t)$ can be taken as $\max_{k\in\mathbb{Z}_+}I(k) \approx \max_{t>0}I(t)$.
Here, for simplicity,
we assume that $S(0)\approx 1$ (and thus $I(0)\approx 0$)
and use the upper bound $I_{\max}(b_0)$ given by
\begin{equation}\label{eq-add01e}
 I_{\max}(b_0)
  = 1  - \frac{1}{b_0 R_0}
     + \frac{1}{b_0 R_0}\ln \frac{1}{b_0 R_0}.
\end{equation}

Note that if \eqref{bR0} does not hold,
i.e., $b_0 R_0 \leq 1$, then the announced
transmission reduction parameter is so small
that $I(k)$ becomes a
nonincreasing function of time and thus
$\max_{k\in\mathbb{Z}_+} I(k) = I(0)\approx 0$.
This is a trivial case
with no infection and hence not of interest
in this paper.

We next introduce a lemma from our previous mobile
malicious work \cite{wang:ifac2020}.
This lemma will be used in the proofs of the
results for the complete graph case studied
in this section. We slightly modified the
statement for the problem setting in this paper.

\begin{lemma} \label{lemma1}
Consider the multi-agent system
under the homogenous epidemic malicious model with the maximum infectious ratio $I_{\max}(b_0)$ in \eqref{eq-add01e},
whose network $\mathcal{G}$ forms a complete graph.
Then, under \eqref{bR0}
the regular agents using Algorithm~\ref{protocol}
reach resilient consensus if and only if
$n > 2\max_{i \in \mathcal{V}}{F_i(k)} + 1
\geq 2I(k)n +1.$
\end{lemma}

\subsection{Protocol for the Static Policy}

When the policy for the transmission reduction parameter
is static, we obtain the following result.
Let $b^*$ be the solution to the equation
\begin{equation} \label{eq-fb}
  2I_{\max}(b^*) = 1 - b^*.
\end{equation}

\begin{proposition}\label{prop01}
Consider the multi-agent system
under the SIR epidemic malicious model whose
network $\mathcal{G}$ forms a complete graph.
If $R_0 >1$,
then the solution $b^*\in(1/R_0,1]$ to \eqref{eq-fb}
always exists. By taking $b_0\in(1/R_0,b^*]$ and
\begin{equation}\label{eqn:Fi0_static}
   F_{i0}\in \Bigl[(1-b_0)\frac{n-1}{2},\frac{n-1}{2}\Big),
\end{equation}
the regular agents using Algorithm~\ref{protocol}
with the static policy
reach resilient consensus.
\end{proposition}

For the complete graph, it holds
$d_i= n-1$ for all $i\in\mathcal{V}$.
Thus, from \eqref{eqn:Fi2},
the choice of $F_{i0}$ becomes
that in \eqref{eqn:Fi0_static}.
For later use, let
\begin{equation}\label{eqn:f}
  f(b) = 2I_{\max}(b) - (1-b).
\end{equation}

\smallskip\noindent
\textit{Proof}:~
We first show that $b^*\in(1/R_0,1]$ satisfying
\eqref{eq-fb} always exists.
Substituting \eqref{eq-add01e} into $f(b)$
we have
\[
 f(b)
  = 1 + b - \frac{2}{b R_0}
     + \frac{2}{b R_0}\ln \frac{1}{b R_0}.
\]
We discuss the property of $f(b)$ for $b\in(1/R_0,1]$.
Since $R_0 > 1$, it is clear that at $b=1/R_0$
\[
f\Bigl(
  \frac{1}{R_0}
 \Bigr)
  = -\Bigl(
       1-\frac{1}{R_0}
     \Bigr) <0.
\]
Then, at $b=1$, it holds
\[
 f(1) = 2 - \frac{2}{R_0} + \frac{2}{R_0} \ln\frac{1}{R_0}
   > 0.
\]
In fact, $f(b)$ is an increasing function
in $(1/R_0,1]$ since
\[
   f'(b)
     = 1 + \frac{2}{b^2R_0} \ln \frac{1}{bR_0} >0.
\]
Thus, it follows that
there always exists
$b^*\in ({1}/{R_0}, 1]$
such that $f(b^*)=0$ and, moreover,
for each $b \in({1}/{R_0}, b^*]$, it holds
$f(b) \le 0$. This indicates that the chosen
$b_0$ satisfies
\[
I_{\max}(b_0) \le \frac{1-b_0}{2} < \frac{1}{2}.
\]
Hence, we have that $F_{i0}$ can be
taken as in \eqref{eqn:Fi0_static}.
From \eqref{eqn:Fi0_static}, it is immediate
that the conditions in Lemma~\ref{lemma1}
are satisfied. As a result,
resilient consensus can be achieved.
\hfill \mbox{$\square$}


\subsection{Protocol for the Dynamic Policy}

Using the dynamic policy for the transmission
reduction $b(k)$, we can adapt it as the epidemic level changes.
The following lemma will be instrumental.
It gives an upper bound on the infectious ratio $I(k)$
under the dynamic policy.

\begin{lemma} \label{lemma03}
Under the epidemic model \eqref{eq-01} \textcolor{black}{with $R_0 >1$},
if the transmission reduction number $b(k)$
takes the dynamic policy \eqref{eqn:bk},
then the maximum infectious ratio meets the relation
\begin{equation}\label{eqn:lemma03}
  I(k)
    < \frac{1}{2}
       \bigg(
          1 - \frac{1}{R_0}
       \bigg)~~\text{for $k\in\mathbb{Z}_+$}.
\end{equation}
\end{lemma}

\medskip\noindent
\textit{Proof}:
By \eqref{eqn:bk}, we have $b(k)\leq 1- 2I(k)$.
It is clear that the infectious ratio
$I(k)$ is smaller if $b(k)$ is larger.
Hence, we take $b(k)$ the largest as $b(k)=1-2I(k)$.
Substituting this into the
epidemic model \eqref{eq-01},
we obtain the dynamics for $I(k)$ as
\begin{equation*}
  I(k+1)
    = I(k) + \beta S(k)I(k)\Delta T \bigl(1-2I(k)\bigr)
       - \gamma I(k)\Delta T. 
\end{equation*}
By the definition of $R_0$, this can be written as
\[
 I(k+1) - I(k)
 = \beta I(k)\Delta T
       \bigg[
          S(k) \bigl(1 - 2I(k)\bigr) - \frac{1}{R_0}
       \bigg].
\]
Since $I(k)$, $\Delta T$, and $\beta$ are
nonnegative,
clearly, the sign of the increment
$I(k+1)-I(k)$ is determined by
\begin{equation}\label{eqn:g}
  g(k) = S(k)\bigl(1 - 2I(k)\bigr) - \frac{1}{R_0}.
\end{equation}
By the assumption $S(0) \approx 1$,
we have $I(0) \approx 0$ and thus,
$g(0) \approx 1- {1}/{R_0} >0$.
From \eqref{eq-01}, if $b(k)>0$,
$S(k)$ is decreasing.
Hence, in the initial period, $I(k)$ is nondecreasing
while $g(k)\geq 0$.
The value of $g(k)$ decreases until
it becomes negative at some time $k_1>0$.
That is, it holds that $g(k)\geq 0$ for
$k\in[0,k_1-1]$ and
$g(k_1)< 0$. Then, by \eqref{eqn:g},
\begin{align*}
  I(k_1) = \max_{k\in[0,k_1]} I(k)
  &\leq \frac{1}{2} -\frac{1}{2R_0 S(k_1)}
  < \frac{1}{2} -\frac{1}{2R_0}.
\end{align*}

Now, we look at $I(k)$ for $k>k_1$.
Suppose that at some time $k_2 > k_1$,
we have $I(k_2) \ge I(k_1)$.
From \eqref{eqn:g}, we have
\begin{align*}
 g(k_2)
  & = S(k_2)\bigl(1 - 2I(k_2)\bigr) - \frac{1}{R_0}
   \le S(k_2)\bigl(1 - 2I(k_1)\bigr) - \frac{1}{R_0} \\
  &\le S(k_1)\bigl(1  -2I(k_1)\bigr) - \frac{1}{R_0}
  =g(k_1)<0.
\end{align*}
Hence, it holds $g(k_2) < 0$ and thus
$I(k_2) < 1/2 -{1}/{(2R_0)}$.
Therefore, for all $k$, we attain
$I(k) < 1/2 -{1}/{(2R_0)}$.
\hfill \mbox{$\square$}

We finally characterize conditions
for resilient consensus.

\begin{proposition}\label{prop02}
Consider the multi-agent system under
the complete network $\mathcal{G}$ where
the malicious agents follow the SIR epidemic
model with $R_0 >1$.
If the pruning number satisfies
$I(k) \cdot n \le F_i(k) <n/2$ for
$i \in \mathcal{S}(k) \cup \mathcal{R}(k)$,
then Algorithm~\ref{protocol} with the dynamic
policy can guarantee
resilient consensus. 
\end{proposition}

\smallskip\noindent
\textit{Proof}:
Lemma~\ref{lemma03} shows in particular that
under the dynamic policy \eqref{eqn:bk} for $b(k)$,
it holds $I(k)<1/2$ at all times.
This is in fact critical for the policy to
maintain $b(k)\geq 0$.
Moreover, the pruning number $F_i(k)$ can
always be selected as in \eqref{eqn:Fi2}.
It is then straightforward to show that the
conditions in Lemma~\ref{lemma1} hold.
As a result, we conclude that resilient
consensus is established.
\hfill \mbox{$\square$}

\subsection{Discussion for the Proposed Two Policies}
\begin{figure}[t]
\centering
\includegraphics[width=1\linewidth]{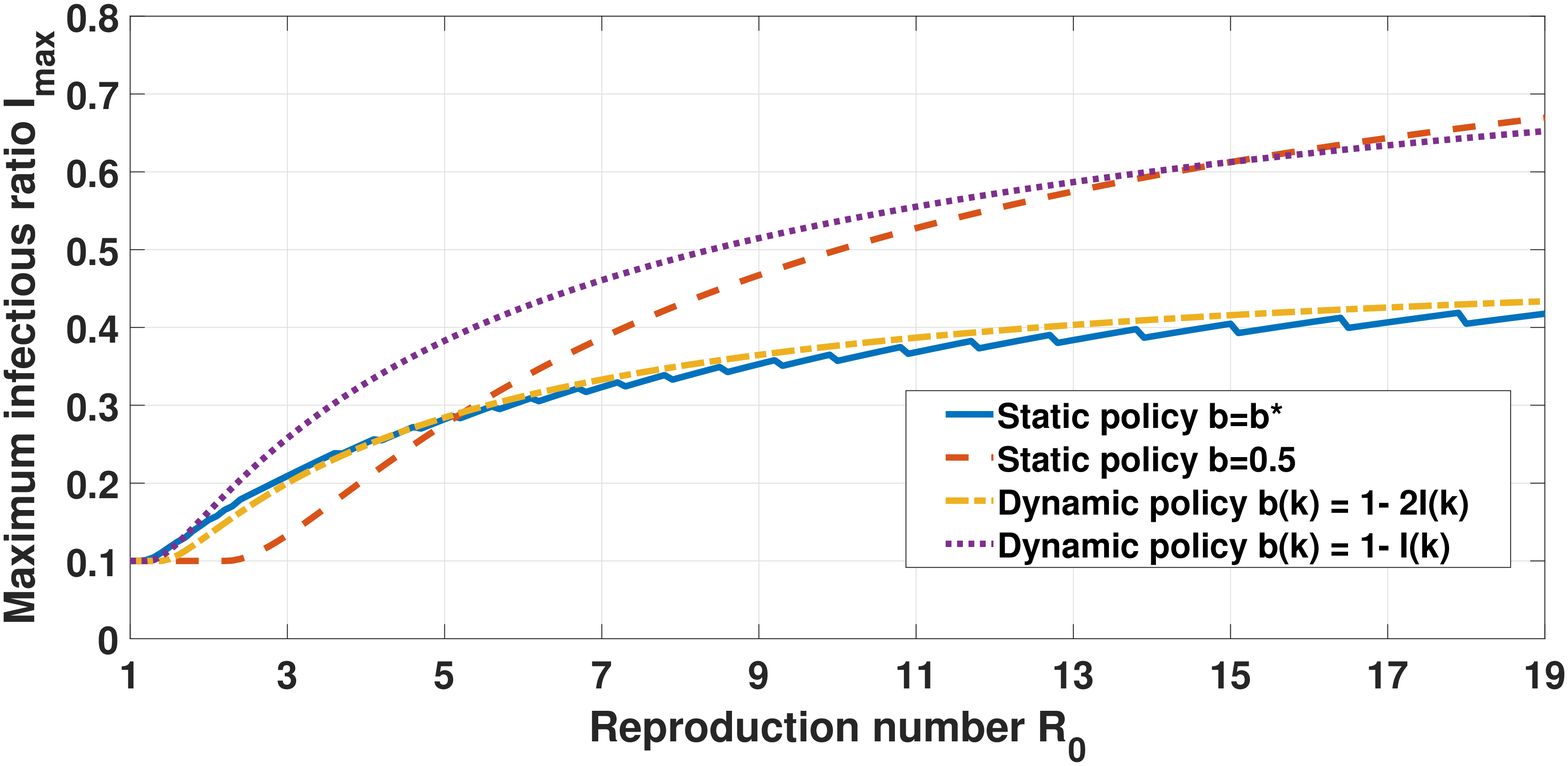}
  \vspace*{-7mm}
\caption{The maximum infectious ratio $I_{\max}$ versus reproduction number $R_0$ under different policies}
\label{fig.add1}
  \vspace*{2mm}
\includegraphics[width=1\linewidth]{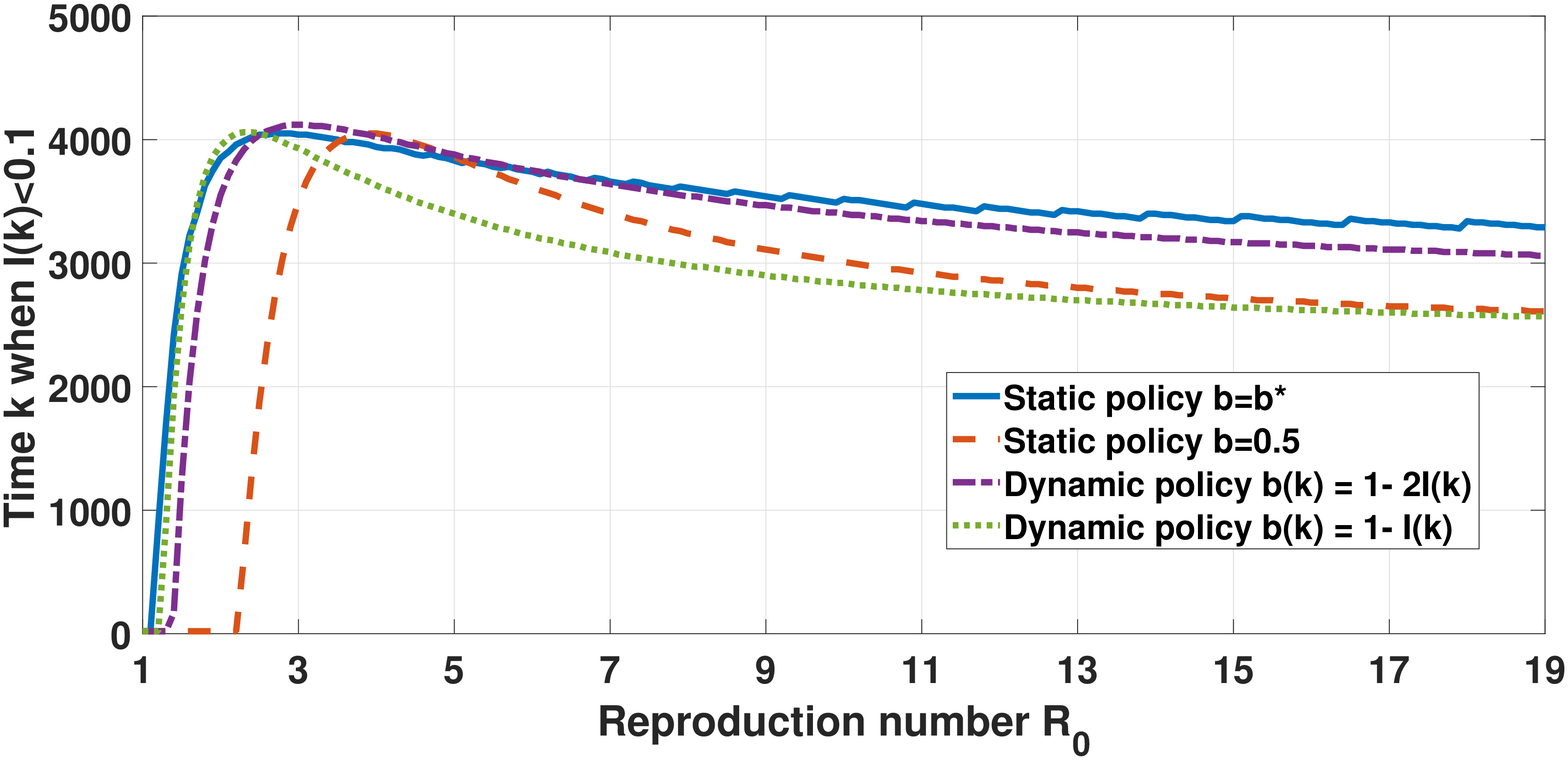}
  \vspace*{-7mm}
\caption{The time when $I(k) < 0.1$ versus reproduction number $R_0$ under different policies}
\label{fig.add2}
  \vspace*{-4mm}
\end{figure}

\vspace{-2 ex}
To compare the two proposed policies, we test them in a range of $R_0 \in [1, 19]$, where initial states are set as $S(0)=0.9, I(0)=0.1, \Delta T=0.01, \gamma =0.1$. Moreover, we change the infectious rate $\beta \in [0.1, 1.9]$ and observe how the maximum infectious ratio changes.
The result can be found in Fig.~\ref{fig.add1}. We can see that both proposed policies can suppress the maximum infectious under $0.5$ for all $R_0 \in [1, 19]$. The other policies such as fixed reduction $b=0.5$ or relaxed dynamic policy $b(k)=1 - I(k)$ cannot guarantee the maximum infectious under $0.5$ when $R_0$ becomes large. This indicates that our proposed policies are well-designed and the infection ratio could be suppressed as a minor part during the whole pandemic process.

Next, to check the approximate
lengths of the pandemic periods under different policies, we show in Fig.~\ref{fig.add2} the time when $I(k) <0.1$ is reached.
From the plots, we can see that for proposed static and dynamic policies, when $R_0 \in [1, 2]$, there is a significant increase in the pandemic period. This indicates that a small $R_0$ may not infect a major part of the agent network. The proposed policies could suppress such weak pandemic processes in a short time with small infectious peaks.
When $R_0 > 3$, the pandemic period does not change too much as $R_0$ increases. The reason is that the pandemic is so powerful that the infectious agents increase rapidly, and then the susceptible agents decrease rapidly so that the pandemic cannot continue for long. With the same recovering rate $\gamma$, they will have similar pandemic periods since the infectious peaks decrease at the same speed.
Note that with fixed reduction $b=0.5$, there is a delay in the increase in the pandemic period. In particular, when $bR_0$ becomes greater than 1, a significant change happens.

\section{Resiliency Results for Noncomplete Graphs}
\label{Section 5}


In this section, we demonstrate that
for agents operating over a noncomplete
graph in the epidemic environment,
resilient consensus can be attained
by the proposed algorithm.
Again, both static and dynamic policies are
considered, and we derive conditions on network structures.

In the noncomplete graph case, the pruning number
$F_i(k)$ for $i\in\mathcal{V}$ should be taken
slightly differently from \eqref{eqn:Fi2} as
\begin{equation} \label{eqn:Fi2_dyn}
 F_{i}(k)
  \in \Bigl[
        \left(1-b_i(k)\right) \frac{n}{2},
        \frac{d_{\min}}{2} - \frac{n}{4}
      \Bigr),
   ~~i\in\mathcal{V}, k\in\mathbb{Z}_+,
\end{equation}
where $d_{\min} = \min_{i\in\mathcal{V}} d_i$.

For the noncomplete graph case,
a sufficient condition on graph structures
is provided in \cite{wang:ifac2020},
which we state as a lemma.

\begin{lemma} \label{lemma2}
Consider the multi-agent system under the SIR
epidemic malicious model.
Then 
the regular agents using Algorithm~\ref{protocol}
with the dynamic policy and \eqref{eqn:Fi2_dyn}
reach resilient consensus
\textcolor{black}{if the following condition holds:
\begin{equation} \label{eqn:lemma2}
  d_{\min}
    > 2\max_{i \in \mathcal{V}}{F_{i}(k)} + \frac{n}{2}
    \ge 2\max_{s \in \{1, \ldots, m \}}{I_{s}(k)} n + \frac{n}{2}.
\end{equation}}
\end{lemma}
\vspace{-6 ex}

\subsection{Protocol for the Static Policy}

We demonstrate the effectiveness of the static policy
for the noncomplete graph case and provide a condition
on the graph structure for achieving resilient
consensus under the epidemic malicious model.

\textcolor{black}{Considering the subgroup heterogeneity, we know that there is at most increment $\overline w$  from the global ratio of infection $I(k)$ to the local ratio of infection $I_s(k)$.
Let $b^*\in(1/R_0,1]$ be the solution to the equation
\begin{equation} \label{eq-fb-heter}
  2\left( I_{\max}(b^*) + \overline w \right) = 1 - b^*.
\end{equation}
Denote $\overline W_s$ as the heterogeneity upper bound for the proposed static policy. We must limit the heterogeneity
by assuming $\overline w < \overline W_s = \frac{1}{2}(1-\frac{1}{R_0})$.
\begin{proposition}\label{prop01-non}
Consider the multi-agent system
under the SIR epidemic malicious model.
If $R_0 >1$ and $\overline w < \overline W_s = \frac{1}{2}(1-\frac{1}{R_0})$,
then the solution $b^*\in(1/R_0,1]$ to \eqref{eq-fb-heter}
always exists.
\end{proposition}
}

\textcolor{black}{
\textit{Proof}:~ The proof follows a similar line as that of Proposition~\ref{prop01}.
Let $f_w(b) = 2I_{\max}(b) + 2\overline w - (1-b)$.
Substituting \eqref{eq-add01e} into this,
we have
\[
 f_w(b)
  = 1 + b + 2\overline w - \frac{2}{b R_0}
     + \frac{2}{b R_0}\ln \frac{1}{b R_0}.
\]
We discuss the property of $f(b)$ for $b\in(1/R_0,1]$.
Since $R_0 > 1$ and $\overline w < \frac{1}{2}(1-\frac{1}{R_0})$, it is clear that at $b=1/R_0$
\[
f_w\Bigl(
  \frac{1}{R_0}
 \Bigr)
  = 2\overline w -\Bigl(
       1-\frac{1}{R_0}
     \Bigr) <0.
\]
Then, at $b=1$, it holds
\[
 f_w(1) = 2 - \frac{2}{R_0} + \frac{2}{R_0} \ln\frac{1}{R_0} + 2\overline w
   > 0.
\]
Furthermore, $f_w(b)$ is an increasing function
in $(1/R_0,1]$ since
\[
   f'_w(b)
     = 1 + \frac{2}{b^2R_0} \ln \frac{1}{bR_0} >0.
\]
Thus, it follows that
there always exists
$b^*\in ({1}/{R_0}, 1]$
such that $f_w(b^*)=0$. \hfill \mbox{$\square$}
}


The design procedure is as follows.
First, each agent~$i\in\mathcal{V}$ must have
sufficiently many neighbors that
\begin{equation}\label{eqn:di_stat_non}
   d_i \in\Big(
            \Big(
               \frac{3}{2}-b^*
            \Big)n,
            n
          \Big).
\end{equation}
Note that by the assumption that $1<R_0<2$, such $d_i$ always exists.
Here, we further assume that
\begin{equation}\label{eqn:f32d}
  f_w\Big(
     \frac{3}{2} - \frac{d_{\min}}{n}
   \Big) < 0.
\end{equation}
Then, take the local transmission reduction $b_{i0}$ satisfying
\begin{equation}\label{eqn:b0_dyn}
   b_{i0} \in\Big(
          \frac{3}{2} - \frac{d_{\min}}{n}, b^*
        \Big).
\end{equation}
The pruning number $F_{i0}$ for agent~$i$ should
be taken as in \eqref{eqn:Fi2_dyn} so that
\begin{equation} \label{eqn:Fi2_stat}
 F_{i0}
  \in \Bigl[
        \left(1-b_{i0}\right) \frac{n}{2},
        \frac{d_{\min}}{2} - \frac{n}{4}
      \Bigr),
   ~~i\in\mathcal{V}.
\end{equation}

Then the following result ensures that the parameters appearing in the procedure above can always be found and they will enable the agents to form consensus.

\begin{theorem}\label{theorem01} 
\textcolor{black}{Consider the multi-agent system under
the network $\mathcal{G}$ with $R_0\in(1,2)$,
where the malicious agents follow the SIR
epidemic model with heterogeneity upper bounded as $\overline w < \overline W_s = \frac{1}{2}(1-\frac{1}{R_0})$}.
Then the regular agents using
Algorithm~\ref{protocol}
with the static policy and
the parameters in
\eqref{eqn:di_stat_non}--\eqref{eqn:Fi2_stat}
reach resilient consensus.
\end{theorem}

\noindent
\textit{Proof}:~
By the choice of $d_i$ in \eqref{eqn:di_stat_non},
we have
${3}/{2} - {d_{\min}}/{n} < b^*$.
Hence, the local transmission reduction parameter $b_{i0}$
satisfying \eqref{eqn:b0_dyn} can be taken. \textcolor{black}{Moreover, since the global transmission reduction parameter $b_0$ is in the interval containing all local transmission reduction parameter $b_{i0}$, we have $b_0 \in (\frac{3}{2} - \frac{d_{\min}}{n}, b^* )$.}
%
Moreover, from \eqref{eqn:b0_dyn}, we have
${1-b_{i0}}/{2}
    < {d_{\min}}/{2n} - {1}/{4}$.
This shows that $F_{i0}$ in \eqref{eqn:Fi2_stat}
is well defined.

\textcolor{black}{Finally, by \eqref{eqn:f32d}, it holds that
$I_{\max}(b_0) + \overline w < {(1-b_0)}/{2}$,
indicating that $\max_{i \in \mathcal{V}}{I_{i}(k)} < {(1-b_0)}/{2}$, and thus $F_{i0}$ from \eqref{eqn:Fi2_stat}
satisfies the condition of Lemma~\ref{lemma2}.
Therefore, resilient consensus can be achieved.}
\hfill \mbox{$\square$}
\vspace{-2ex}

\subsection{Protocol for the Dynamic Policy}

\textcolor{black}{
Now, we give the result when the dynamic
policy is used for the noncomplete graphs case.
Denote by $\overline W_d$ the heterogeneity upper bound for the proposed dynamic policy.
We must limit the level of adversarial spreading and heterogeneity
by assuming $R_0\in(1,2)$ and $\overline w < \overline W_d = \frac{1}{2R_0}-\frac{1}{4}$.
Then, each agent~$i\in\mathcal{V}$ takes
enough neighbors so that
\begin{equation}\label{eqn:di_dyn_non}
   d_i \in\Big(
            \Big(
               \frac{3}{2} + 2 \overline w -\frac{1}{R_0}
            \Big)n,
            n
          \Big).
\end{equation}
Here, due to the condition on $R_0$ and $\overline w$,
it holds $3/2-1/R_0 + 2\overline w \in(1/2,1)$ and
hence, by \eqref{eqn:di_dyn_non}, we have $d_i> n/2$.}

\begin{theorem}\label{theorem03}
Consider the multi-agent system under
the network $\mathcal{G}$ where the
malicious agents follow the SIR epidemic
model with $R_0\in(1,2)$, \textcolor{black}{where the malicious agents follow the SIR
epidemic model with heterogeneity bounded from above by $\overline w < \overline W_d = \frac{1}{2R_0}-\frac{1}{4}$.}
Then the regular agents using
Algorithm~\ref{protocol} with the dynamic
policy \eqref{eqn:bk} using parameters in \eqref{eqn:Fi2_dyn} and
\eqref{eqn:di_dyn_non}
reach resilient consensus.
\end{theorem}

\textit{Proof}:~
To establish resilient consensus,
we must show that the condition
in Lemma~\ref{lemma2} holds.
\textcolor{black}{Under the dynamic policy, by Lemma~\ref{lemma03},
it holds $2 I(k) < 1 - {1}/{R_0}$
at all times. Hence, by \eqref{eqn:di_dyn_non},
\begin{flalign}
  d_i &> \Big(1-\frac{1}{R_0}\Big)n + 2\overline w n +\frac{1}{2}n \nonumber \\
      &> 2 I(k)n + 2\overline w n + \frac{1}{2}n \nonumber \\
      &> 2\max_{s \in \{ 1,\ldots,m \}}{I_{s}(k)}n + \frac{1}{2}n. \nonumber
\end{flalign}
Thus, the part of the condition \eqref{eqn:lemma2}
in Lemma~\ref{lemma2} holds.
By the choice of
$F_i(k)$ in \eqref{eqn:Fi2_dyn} and the dynamic
policy \eqref{eqn:bk},
it follows that
\eqref{eqn:lemma2} is satisfied.}
\hfill \mbox{$\square$}

Note that this result is quite conservative
since Lemmas~\ref{lemma03} and~\ref{lemma2}
have conservatisms.
The actual bound for $R_0$ may be much more
relaxed. We will discuss this point
by simulations in the next section.
\vspace{-1ex}

\section{Numerical Example}
\label{Section 6}

We illustrate the performance of our proposed
protocols under epidemic adversary models by
a numerical example.

Networks with 1000 nodes were generated
by randomly placing
nodes having the communication radius of $r$
in the area of 100 $\times$ 100.
The connectivity requirements are in general
difficult to check. \textcolor{black}{There are two local policy makers placed in the network, and thus $s=2$. Each subgroup contains 500 nodes, and the local ratio of the infection $I_s(k)$ is available for the policy maker.}
The initial number of infected agents is set to 10 so that $I(0)=0.01$,
$S(0)=0.99$, and $R(0)=0$.
To ensure the cardinality of the sets
$\mathcal{S}(k)$, $\mathcal{I}(k)$, and $\mathcal{R}(k)$
to be integers, we took
$|\mathcal{S}(k)| = \lceil S(k) \cdot n \rceil$,
$|\mathcal{R}(k)| = \lceil R(k) \cdot n \rceil$, and
$|\mathcal{I}(k)| = n- |\mathcal{S}(k)| - |\mathcal{R}(k)|$.
For simplicity, all agents use the same pruning number,
denoted by $F$. To check the success rate of resilient consensus under different conditions, Monte Carlo simulations were made. Initial states of the non-infectious agents were randomly taken in the interval of $[0,1]$. On the other hand, the infected agents were forced to take negative state values at $-1$.
\vspace{-2ex}

\subsection{Complete Graph Case}

To build complete graphs, we set the communication radius to be
large ($r=150$). 
Based on Proposition~\ref{prop01}, we know
that for the reproduction number satisfying $R_0 > 1$, we can always find $b_0 \in[1/R_0,b^*)$
so that we can choose $ F\in((1-b_0)(n-1)/{2},(n-1)/{2})$
to guarantee resilient consensus.
To verify this result, we chose $R_0=200$ and $F=499$
and ran 50 Monte-Carlo simulations.
In all cases, resilient consensus was achieved.
For the dynamic policy, we took $F(k) = \lceil I(k)\cdot n \rceil$;
the results also confirmed successful resilient consensus
in all 50 simulations.

\vspace{-2 ex}
\subsection{Noncomplete Graph Case}

For non-complete graphs, we examined how the
parameters $R_0$, $r$, and $F$ affect resilient
consensus under the proposed algorithm. \textcolor{black}{In this subsection, we test the homogeneous adversarial spreading and thus $\overline w =0$. Hence, the heterogeneity requirements $\overline w < \overline W_s$ and $\overline w < \overline W_d$ in Theorems~\ref{theorem01} and~\ref{theorem03} are both satisfied. We demonstrate the effects of infectious heterogeneity in the next subsection.}
First, we focus on the resilience resulting
from the network structure. To this end,
the communication radii were chosen as $r=90$ and $r=50$,
which represent dense and sparse networks, respectively.
For each pair $(R_0,F)\in[1,3]\times[10,350]$
of the reproduction number and the pruning number,
we performed 50 Monte Carlo simulations to find
the success rates for resilient consensus.

\begin{figure}[t]
  \centering
  \includegraphics[width=1\linewidth]{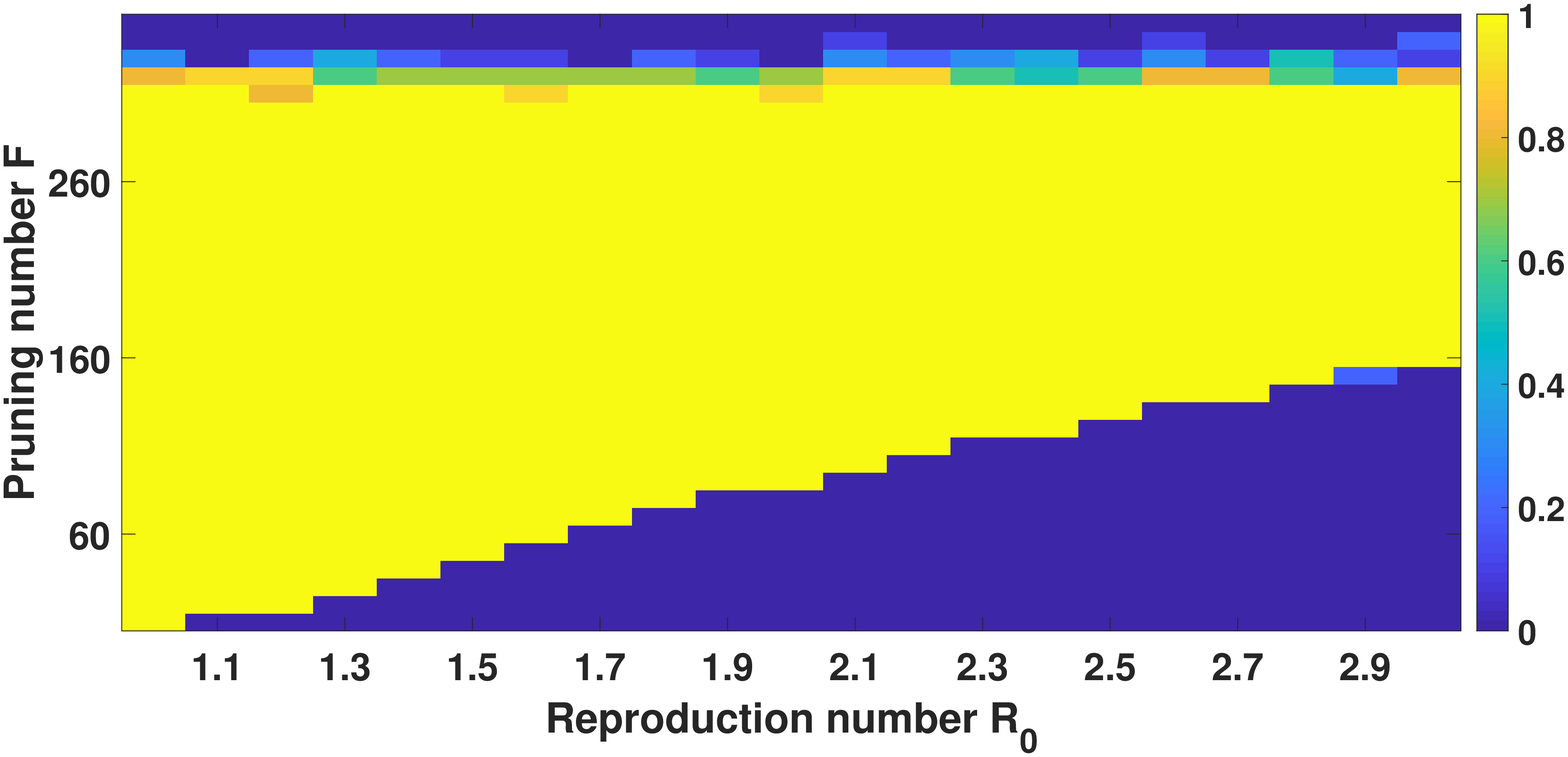}
  \vspace*{-7mm}
  \caption{Static policy: Success rates for resilient
consensus versus reproduction number $R_0$
and pruning number $F_{i0}$ with parameter $r=90$.}
  \vspace*{2mm}
  \label{fig3}
  \includegraphics[width=1\linewidth]{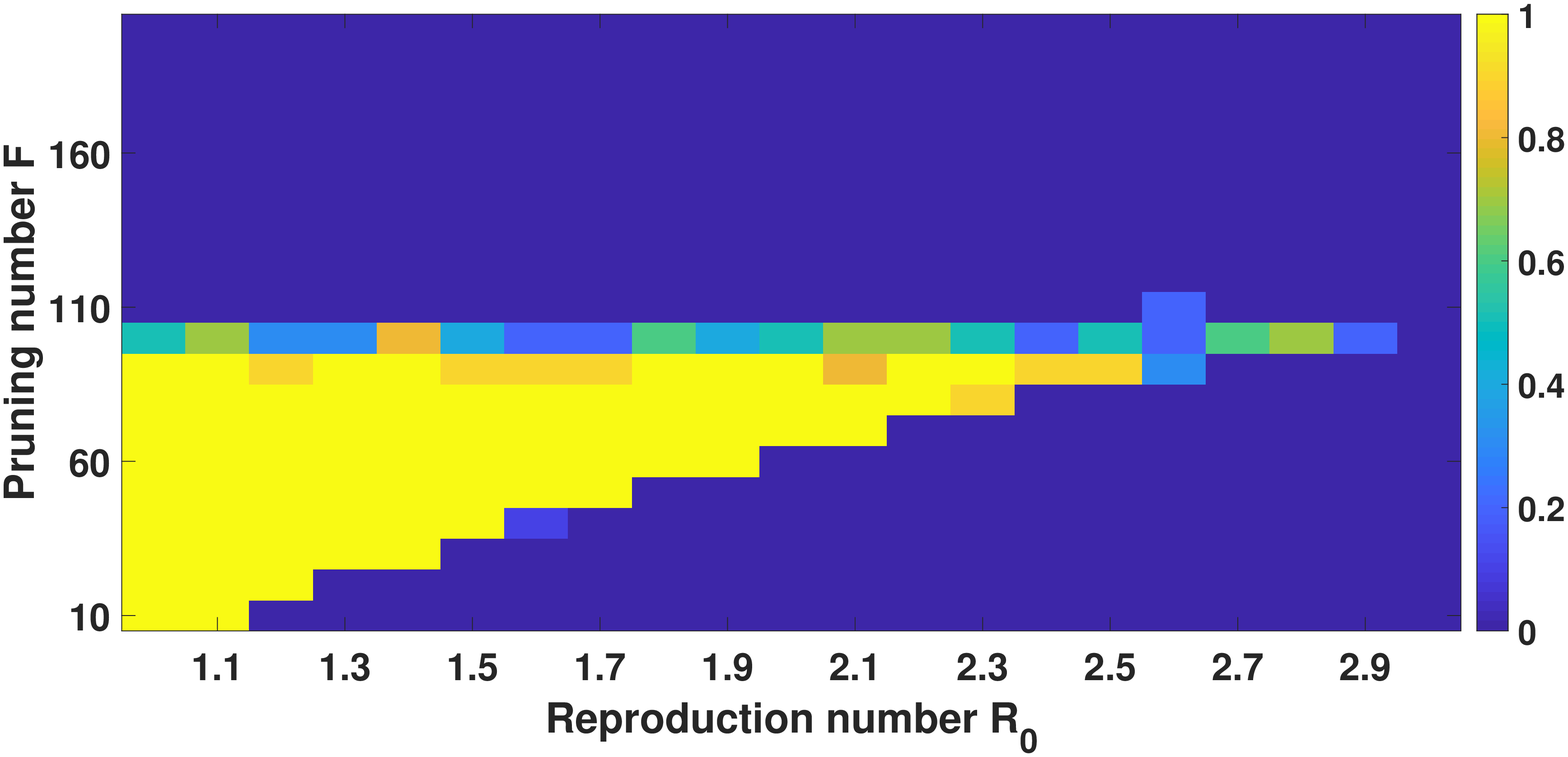}
  \vspace*{-7mm}
  \caption{Static policy: Success rates for resilient
consensus versus reproduction number $R_0$
and pruning number $F_{i0}$ with parameter $r=50$.}
  \vspace*{-4mm}
\label{fig4}
\end{figure}

The results are shown in Figs.~\ref{fig3} and~\ref{fig4} in
the form of heatmaps, where colors varying from yellow to
blue indicate the success rates from 1 to 0.
Fig.~\ref{fig3} shows that in dense networks (with $r=90$),
it is possible to achieve resilient consensus even
when the reproduction number $R_0$ is as large as 3.
As the epidemic becomes more powerful with larger $R_0$,
the yellow area shrinks; the lower bound on $F$
increases while its upper bound remains about the same.
The upper bound of $F$ is determined by the
graph connectivity; when a large pruning number
(such as $F > 320$) is used,
the network will become too sparse for the MSR
to perform properly,
leading to failure in reaching consensus.

From Fig.~\ref{fig4}, we notice that
in sparse networks (with $r=50$),
the yellow area becomes limited in size compared
with dense networks (with $r=90$). (Note however
that the scales in $y$-axis of the plots are slightly different, and
the former is not a subset of the latter.)
In particular, resilient consensus is almost
impossible for $R_0 > 2.9$. The upper bound
of pruning number $F$ for this graph is also
much smaller, around 100.
Therefore, to summarize, with more connectivity in the network,
the multi-agent system can tolerate more powerful epidemics.
For sparse networks, resilient consensus may be
hard to guarantee and a feasible transmission
reduction parameter may not exist. These observations
are in alignment with the discussions related to
Theorem~\ref{theorem01}.

\begin{figure}[t]
\centering
\includegraphics[width=1\linewidth]{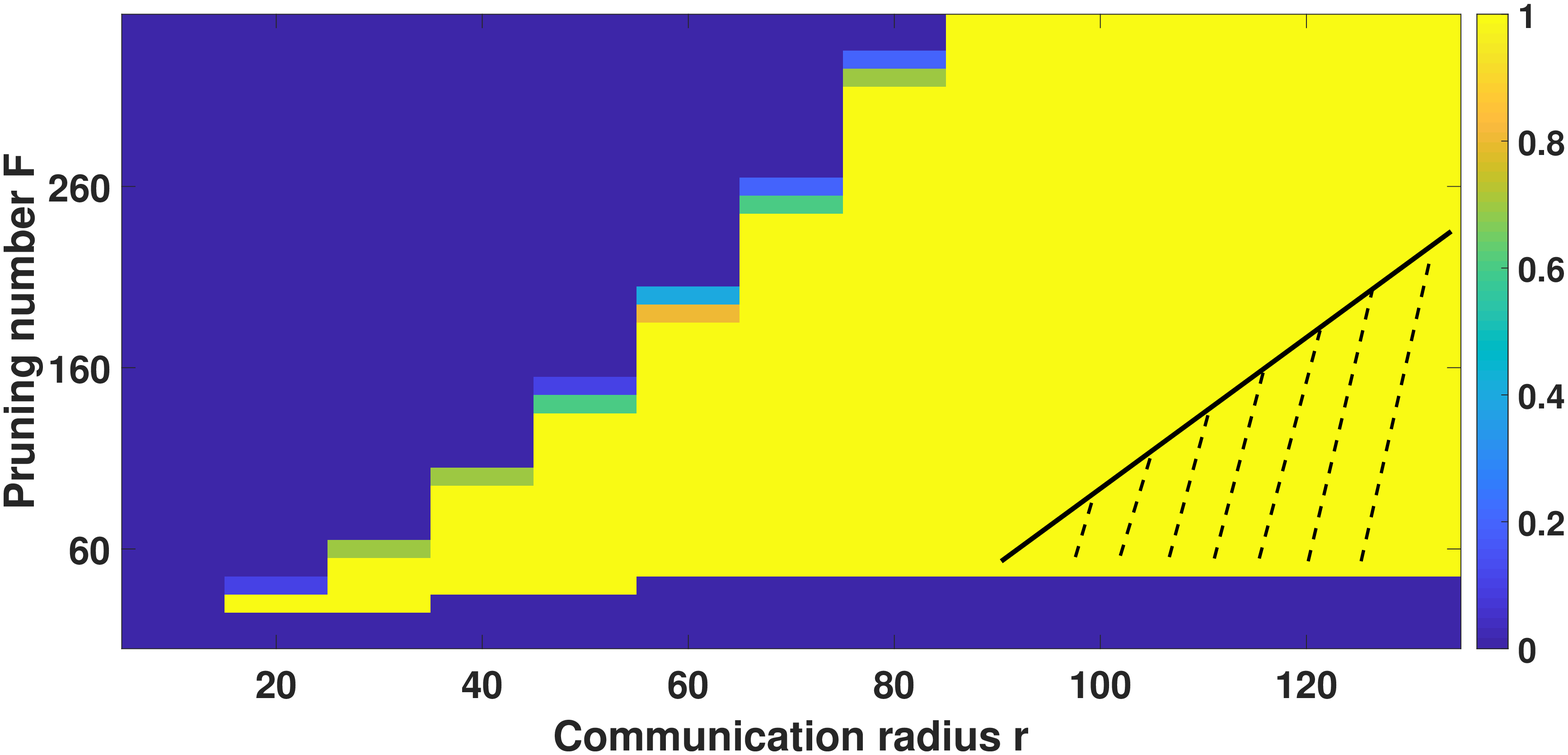}
  \vspace*{-7mm}
\caption{Static policy: Success rates for resilient
consensus versus communication radius $r$ and
pruning number $F_{i0}$
with parameter $R_0=1.5$.
Shaded area: Approximate pruning number from \eqref{eqn:Fi2_stat}.}
\label{fig5}
  \vspace*{2mm}
\includegraphics[width=1\linewidth]{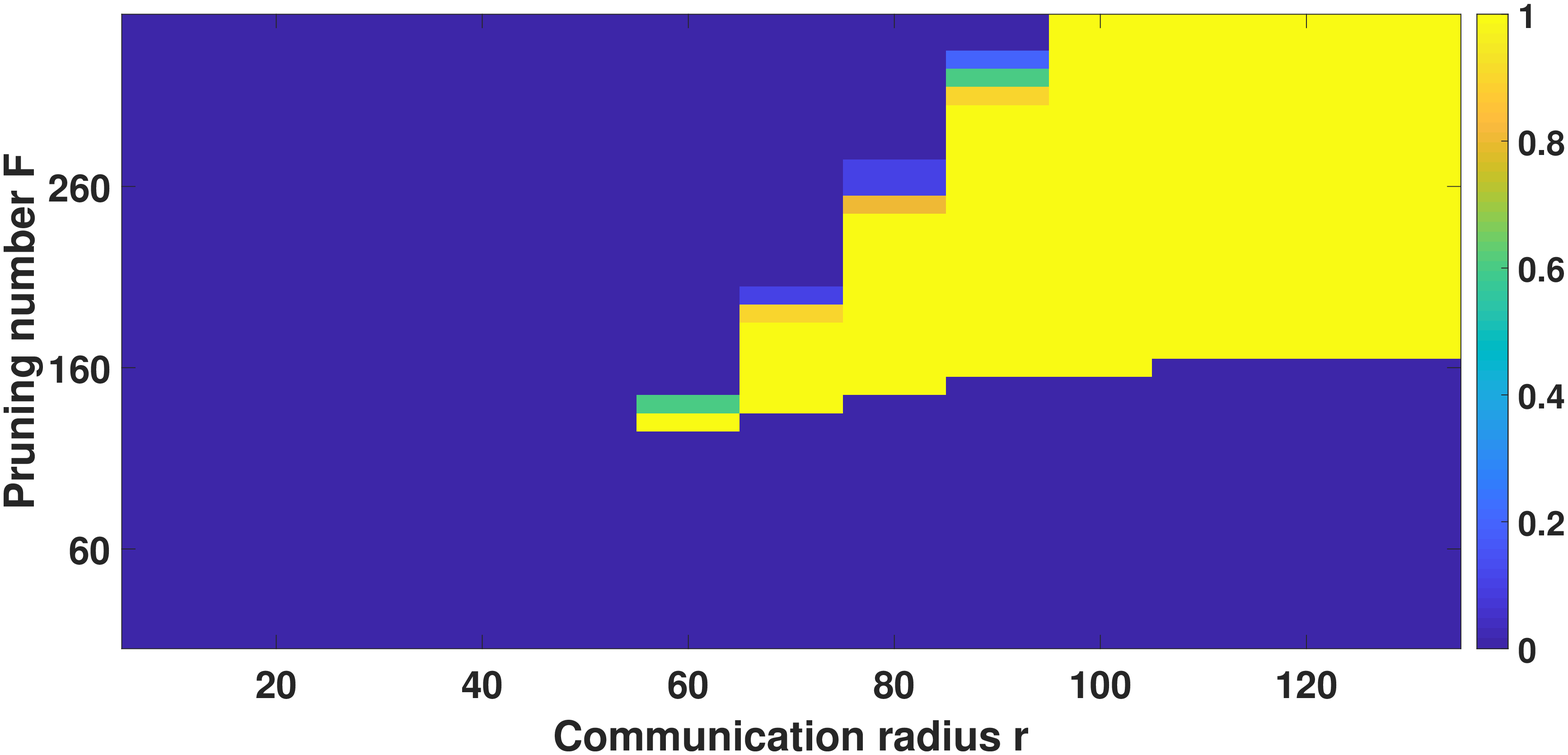}
  \vspace*{-7mm}
\caption{Static policy: Success rates for resilient
consensus versus communication radius $r$ and
pruning number $F_{i0}$ with parameter $R_0=3$.}
\label{fig6}
  \vspace*{-4mm}
\end{figure}

Next, we slightly change our viewpoint and look
at the effects of strength in the adversarial spreading
through the reproduction number $R_0$ on the performance
of resilient consensus algorithms.
Here, we set $R_0=1.5$ for the weaker case and $R_0=3$
for the stronger case.
The success rates for resilient consensus are
computed for the pairs $(r,F)\in[10,130]\times [10,350]$
of the communication radius $r$ and the pruning number $F$.
Again in the form of heatmaps,
the results are shown in Figs.~\ref{fig5} and~\ref{fig6}
for $R_0=1.5$ and $R_0=3$, respectively.

For the results of $R_0=1.5$ in Fig.~\ref{fig5},
it is demonstrated that the minimum of the pruning number $F$
is about $50$ while the maximum increases for denser networks
with larger radii $r \ge 40$.
This indicates that in the simulations,
the number of infectious agents
grew from the initial number of 10 to 50.
The shaded area in the plot indicates
the bound obtained from \eqref{eqn:di_dyn_non},
exhibiting its conservatism;
it shows the necessary radius $r$ to meet the
formula using $d_{\min}$.

From the simulation results, if the communication radius
reaches $r=40$, we can almost guarantee the success rate
to be 1 if $F=50$. However, observe the minimum degree
of each agent in this graph,
it shows that it is smaller than $n/2$.
From the theoretical bound of this paper,
we need the minimum connection of each agent should
reach $600$. We confirmed that for $r \ge 90$,
this condition was guaranteed in the 50 runs.
This indicates the difference between the actual bound
and the theoretical bound.

Under the stronger epidemics with $R_0 = 3$, we can
make similar observations from Fig.~\ref{fig6}.
However, clearly more connectivity is required
so that larger values of $F$ can be used
to guarantee resilient consensus.
The minimum requirements are roughly
$r \ge 70$ and $F\geq 150$.

\begin{figure}[t]
\centering
\includegraphics[width=1\linewidth]{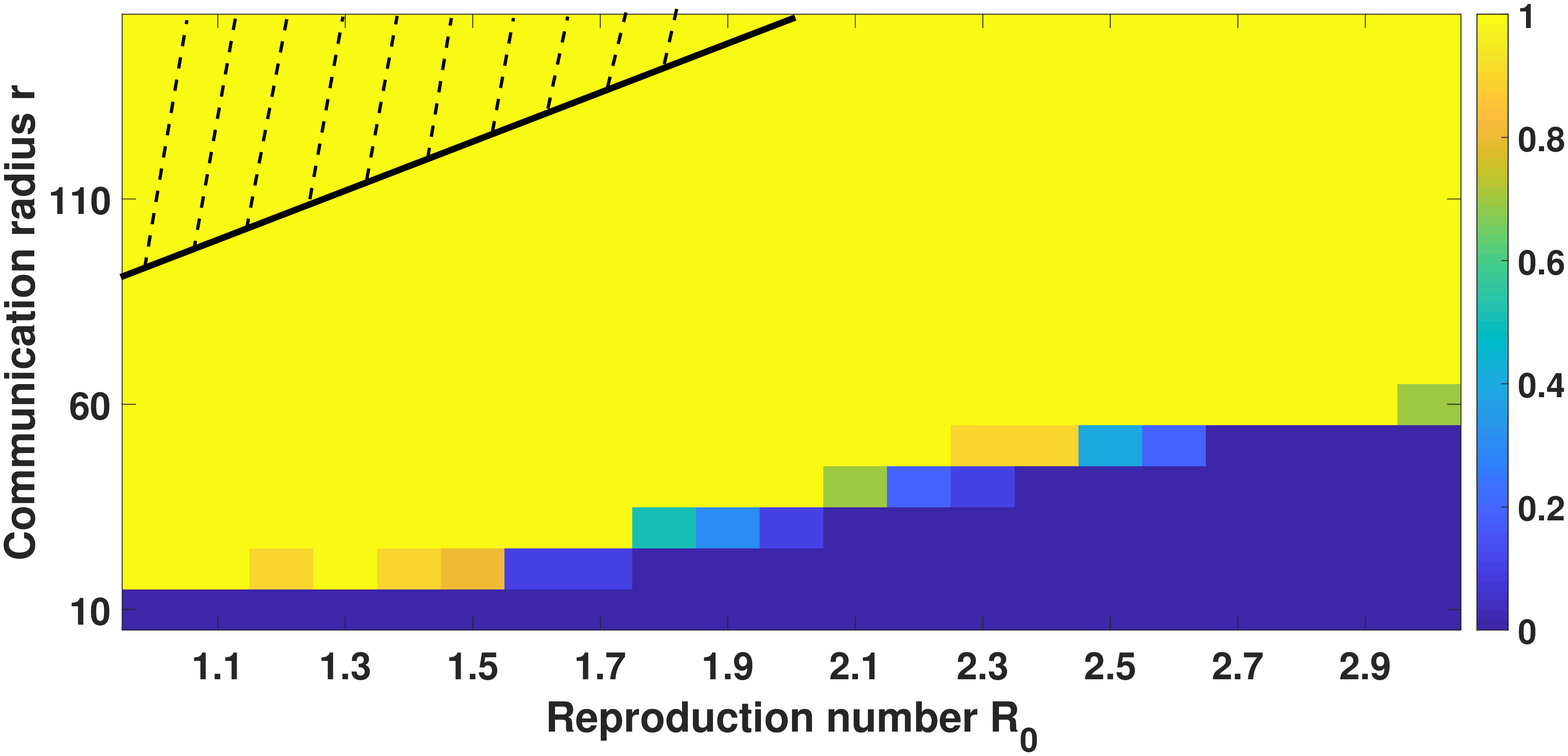}
  \vspace*{-7mm}
\caption{Dynamic policy: Success rates for resilient
consensus versus reproduction number $R_0$
and communication radius $r$.
Shaded area: Approximate connection requirement
from \eqref{eqn:di_dyn_non}.}
\label{fig7}
  \vspace*{2mm}
\includegraphics[width=1\linewidth]{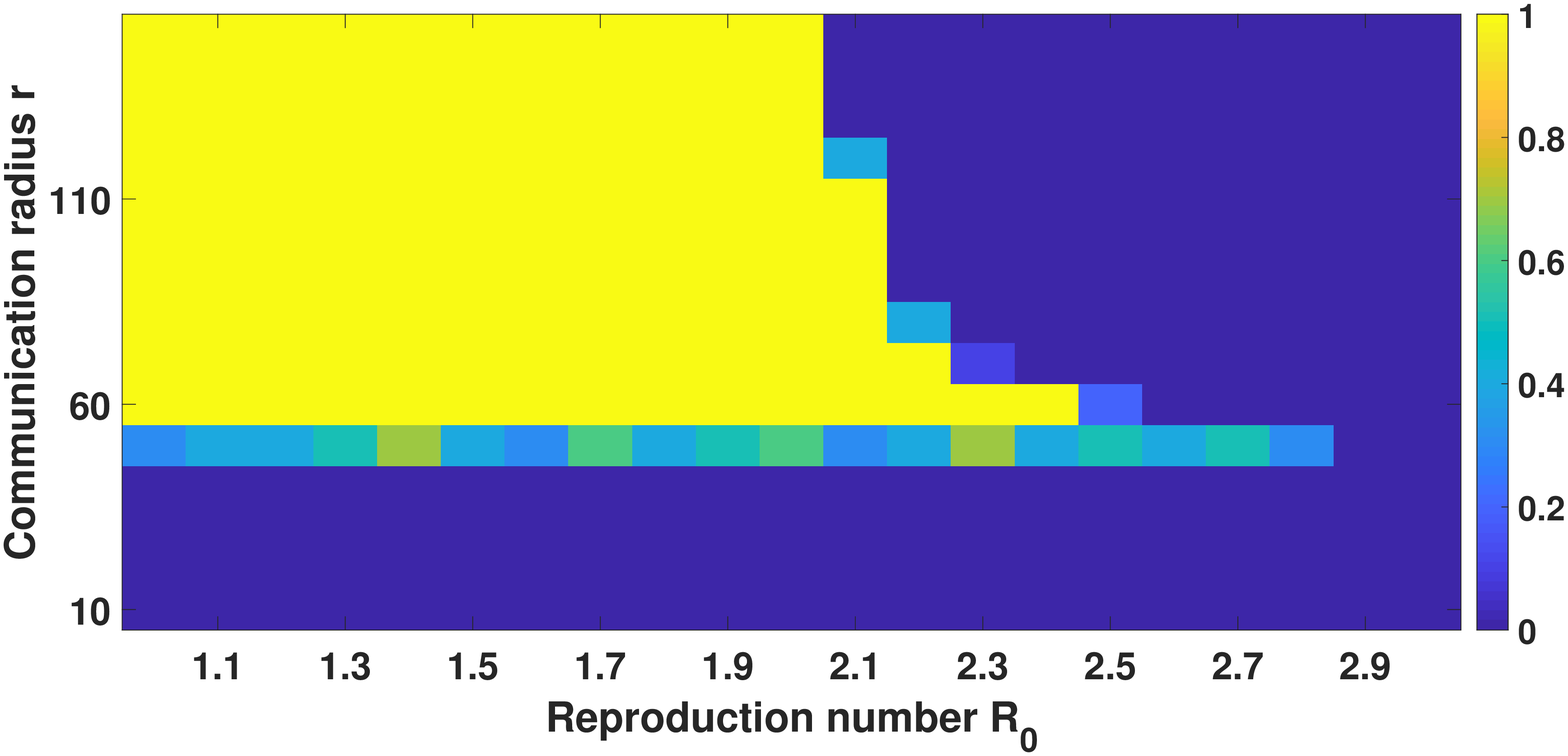}
  \vspace*{-7mm}
\caption{Static policy without policy maker: Success rates for resilient
consensus versus reproduction number $R_0$ and
communication radius $r$ with $F_{i0}=100$.}
\label{fig8}
  \vspace*{-4mm}
\end{figure}


Here, we discuss the performance of
the proposed algorithms with dynamic $F_i(k)$.
Comparisons are made with the static approach
using $F_{i0}=100$.
The heatmaps representing the success rates of resilient
consensus 
are displayed in Figs.~\ref{fig7} and~\ref{fig8}
for $(R_0,r)\in[1,3]\times[10,150]$.

In comparison, the dynamic policy works well
under a range of conditions. This is because the regular
agents can adapt their pruning number based on the current
infectious level as $F_i(k) \ge I(k)\cdot n$
guaranteeing $F_i(k)$ to be always larger than the actual
number of infectious agents in realtime.
In Fig.~\ref{fig7}, observe that as $R_0$ increases,
by choosing denser networks (with larger $r$),
the adaptive rule attains resilient consensus.
Note that our theoretical result in Theorem~\ref{theorem03} for the
dynamic policy
guarantees resilient consensus in dense
networks with limited $R_0$,
located in the shaded area at the left top of Fig.~\ref{fig7}.

Fig.~\ref{fig8} exhibits the results for the static
policy without policy maker where
the pruning number is set as $F_{i0}=100$.
It is clear by comparing this plot with Fig.~\ref{fig7}
that this protocol is much less capable
because of the fixed pruning number.
The minimum requirement on the radius is $r=60$
since each agent must remove 200 values from it neighbors.
The advantage of the static policy is that the agents
do not need the realtime information of $b(k)$.
Under mild epidemics with $R_0 < 2$,
the agents may choose the communication radius $r \ge 60$.
Then, with $F_{i0}=100$, resilient consensus can be guaranteed.

\vspace{-1 ex}
\subsection{\textcolor{black}{The Effects of Heterogeneity}}
In this section, we would like to highlight the effects of the \textcolor{black}{subgroup heterogeneity} used in the analysis regarding the spreading of adversaries in the network.
\textcolor{black}{We test both static and dynamic policies in the homogeneous infected environment and in the gathered infected environment. Note that in the gathered infected environment, the infection spread primarily in subgroup 2. Only after all the 500 agents in subgroup 2 are infected or recovered, the newly infected nodes will appear in subgroup 1. Intuitively, the local infection ratio in subgroup 2 will be much higher than the global infection ratio $I(k)$. Thus, if the network is sparse (i.e., when the communication radius $r$ is small), then the heterogeneity, which may be measured by the size of $\overline w$, could be large, and in the simulations, it is possible that $\overline w > \overline W_s$ or $\overline w > \overline W_d$.}

\begin{figure*}[t]
  \centering
  \subfigure[Static policy within homogeneous infection environment]{%
      \label{fig.new1}
      \includegraphics[width=0.48\linewidth]{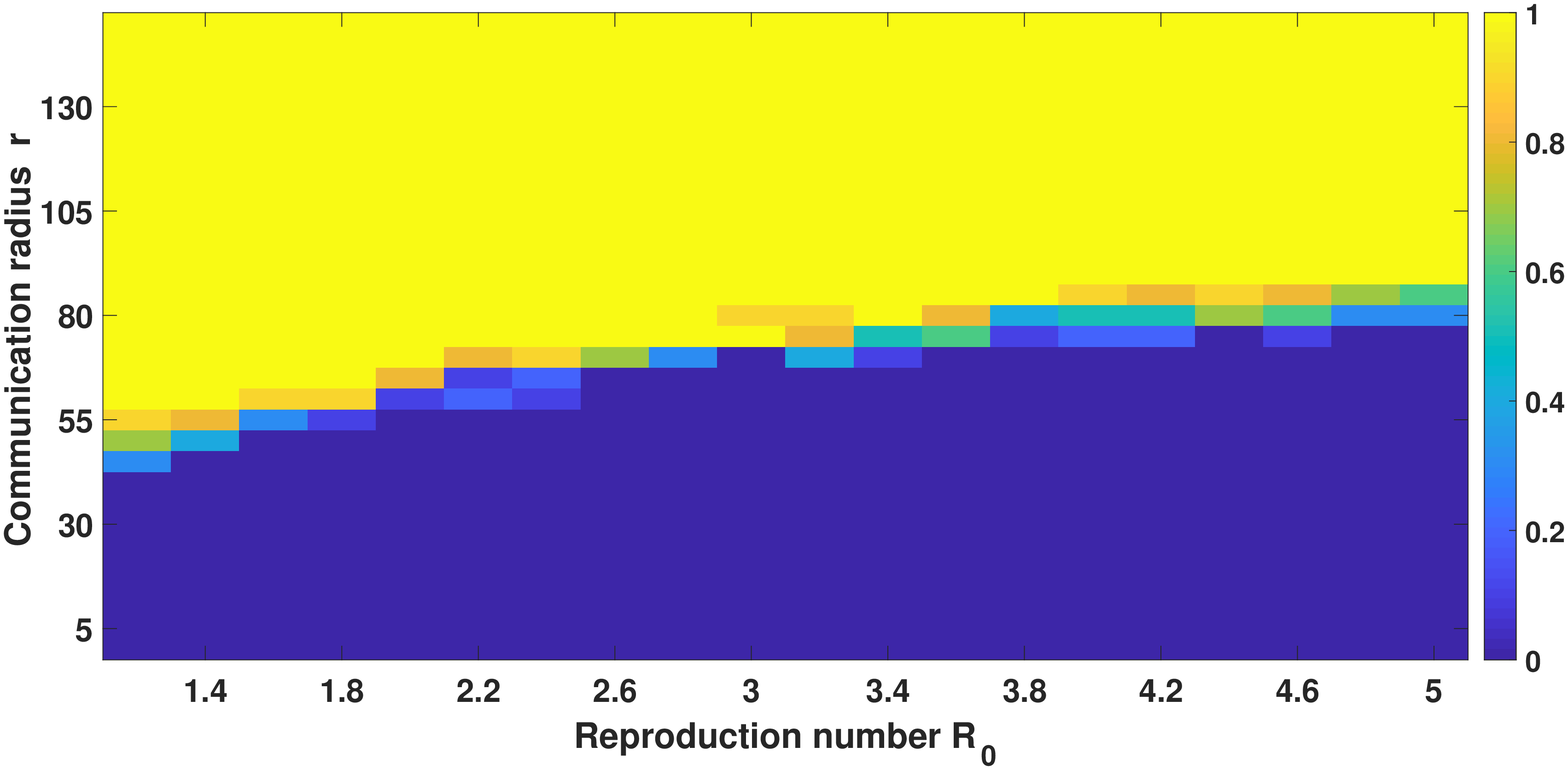}
      \vspace*{-2cm}}
  \subfigure[Static policy within gathered infection environment]{%
      \label{fig.new3}
      \includegraphics[width=0.48\linewidth]{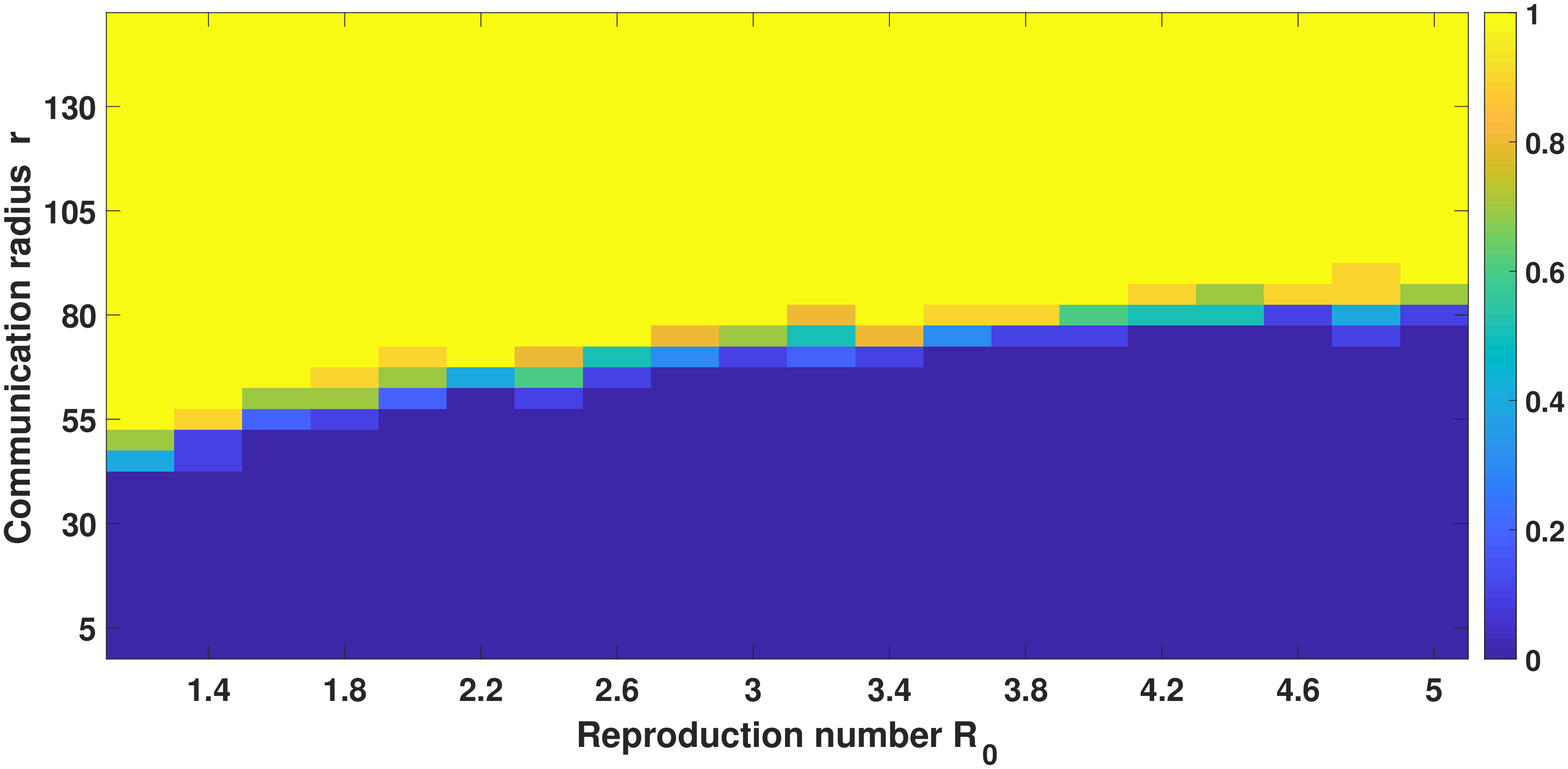}
      \vspace*{-2cm}}
  \subfigure[Dynamic policy within homogeneous infection environment]{%
      \label{fig.new5}
      \includegraphics[width=0.48\linewidth]{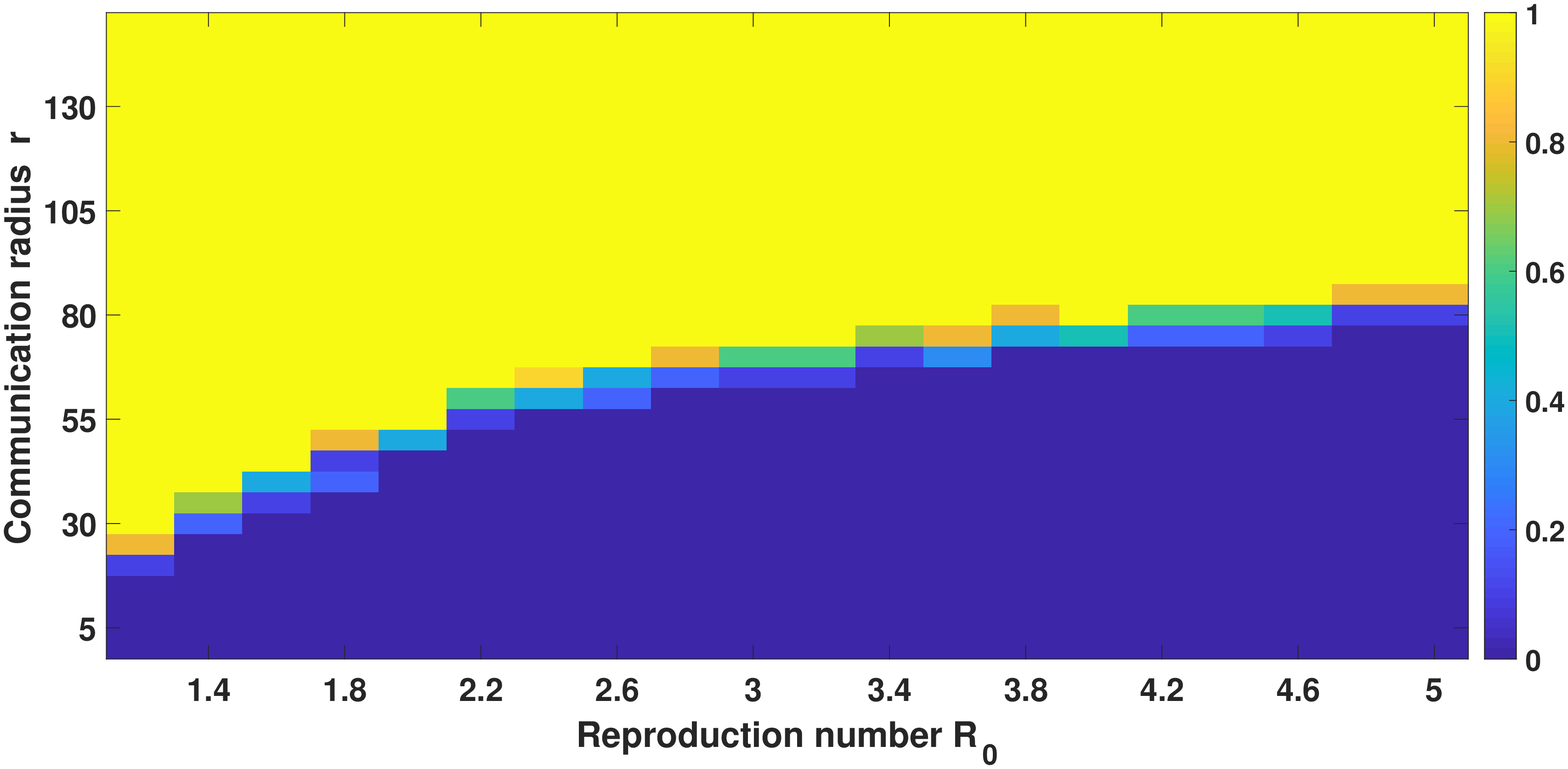}
      \vspace*{-2cm}}
  \subfigure[Dynamic policy within gathered infection environment]{%
      \label{fig.new6}
      \includegraphics[width=0.48\linewidth]{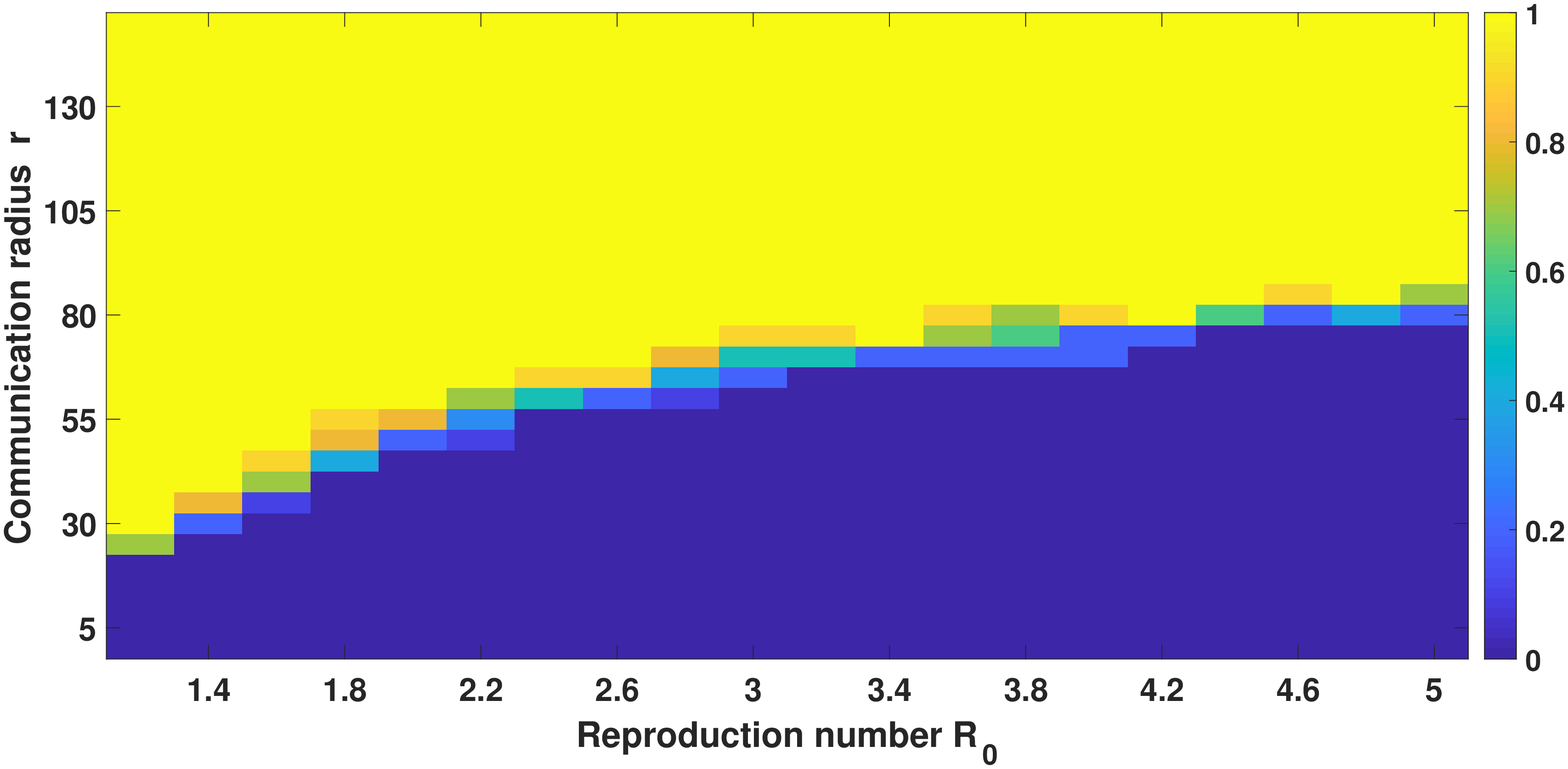}
      \vspace*{-2cm}}
  \caption{Success rates of resilient consensus
           versus the  $R_0$ of SIR model
           and $r$ in the regular agents}
 \label{fig.ggg}
\end{figure*}

\textcolor{black}{We test the proposed static policy without the theoretical connectivity conditions given in \eqref{eqn:di_stat_non}--\eqref{eqn:Fi2_stat}. Instead, in each subgroup $s \in \{1,2\}$, the local policy makers choose $b_0=b^*$ and the regular agents choose $F_{i0} = \lceil (1-b_0)\frac{n}{2} \rceil$. On the other hand, for the dynamic policy, the regular agents in each subgroup $s$ follow the local transmission reduction parameter $b_s(k)$ as in~\eqref{eqn:bk} and set their pruning number as $F_{i}(k) = \lceil (1-b_s(k))\frac{n}{2} \rceil$.
The condition in \eqref{eqn:di_dyn_non} may not be met as $\overline{w}$ may be too large.
We run these two policies in both homogeneous infected environment and gathered infected environment.
The heatmaps representing the success rates of resilient consensus for four different cases are displayed in Fig.~\ref{fig.ggg} for $(R_0,r)\in[1,5]\times[0,150]$.}

\textcolor{black}{We first discuss the results for the static policy. As shown by the yellow regions in Figs.~\ref{fig.new1} and \ref{fig.new3}, for dense networks ($r>100$), the static policy works for almost any $R_0 \in [1,5]$.
This is in alignment with what we have shown in Section~\ref{Section 4}.
For sparse networks ($r < 50$), we can see that the static policy works with limited $R_0$.
The results also show that the proposed static policy works similarly in both homogeneous and gathered infected environments.
The reason is that we chose $F_{i0} = \lceil (1-b_0)\frac{n}{2} \rceil$, which is based on the worst-case analysis. When the network connection is satisfied so that all agents have enough neighbors, even the gathered infectious distribution is a minor part so that all infected agents can be removed in the MSR algorithm as $F_{i0} \ge (I_{\max} + \overline w)n > \left|
       \mathcal{N}_i \cap\mathcal{I}(k)
  \right|$. }

\textcolor{black}{Similar explanations also apply to the dynamic policy whose
results are shown in Figs.~\ref{fig.new5} and~\ref{fig.new6}.
However, note that under small $R_0$, this
policy performs better than the static one especially
when the network is sparse ($r<60$).
Since the policy can dynamically adjust the
pruning number in the MSR to the situation in
real time, even with fewer neighbors, the policy
can manage to reach resilient consensus under a wider range of conditions as expected.
}

\begin{figure*}[t]
  \centering
  \subfigure[Local static policy within homogeneous infection environment]{%
      \label{fig.new7}
      \includegraphics[width=0.48\linewidth]{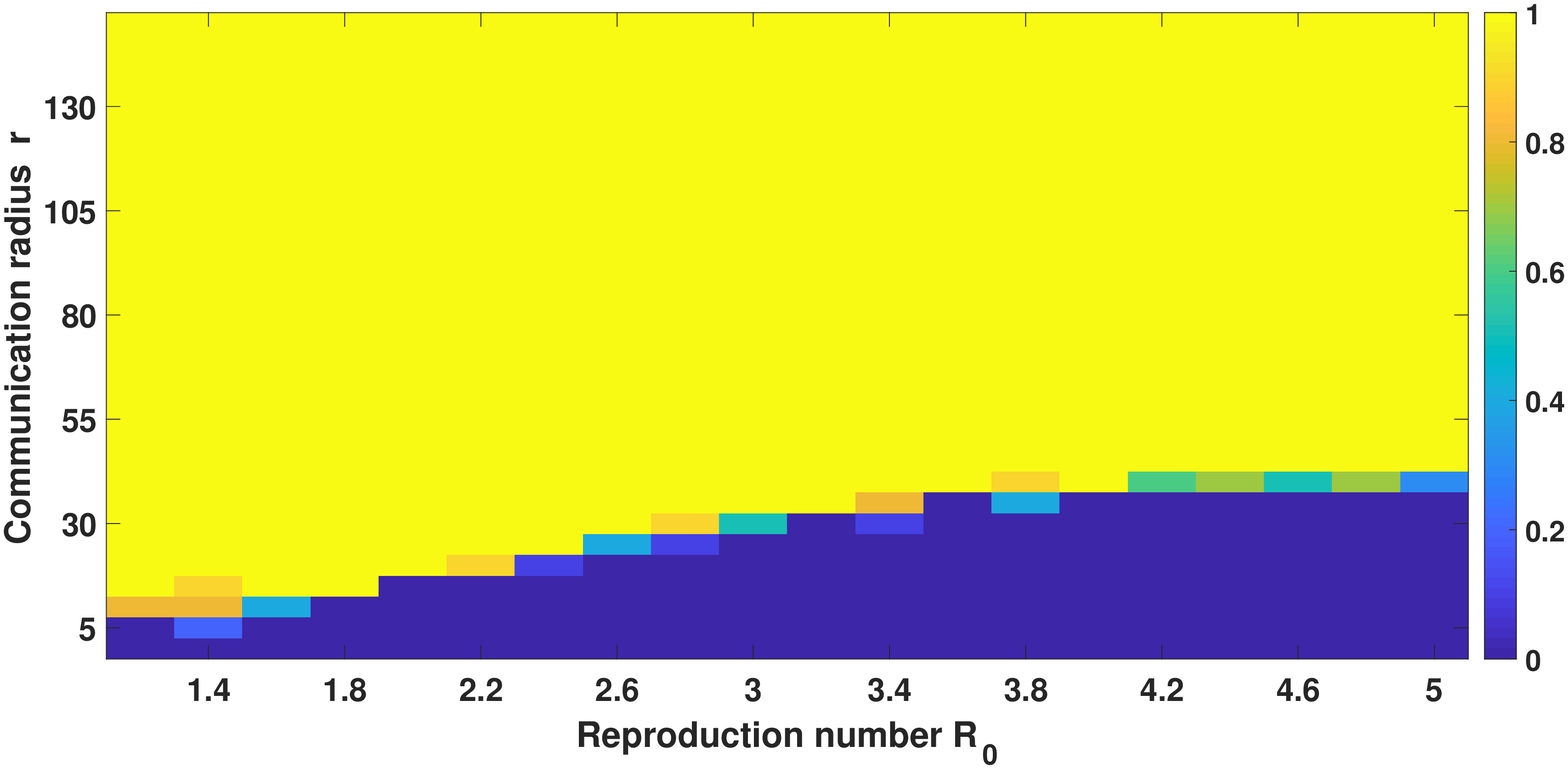}
      \vspace*{-2cm}}
  \subfigure[Local static policy within gathered infection environment]{%
      \label{fig.new8}
      \includegraphics[width=0.48\linewidth]{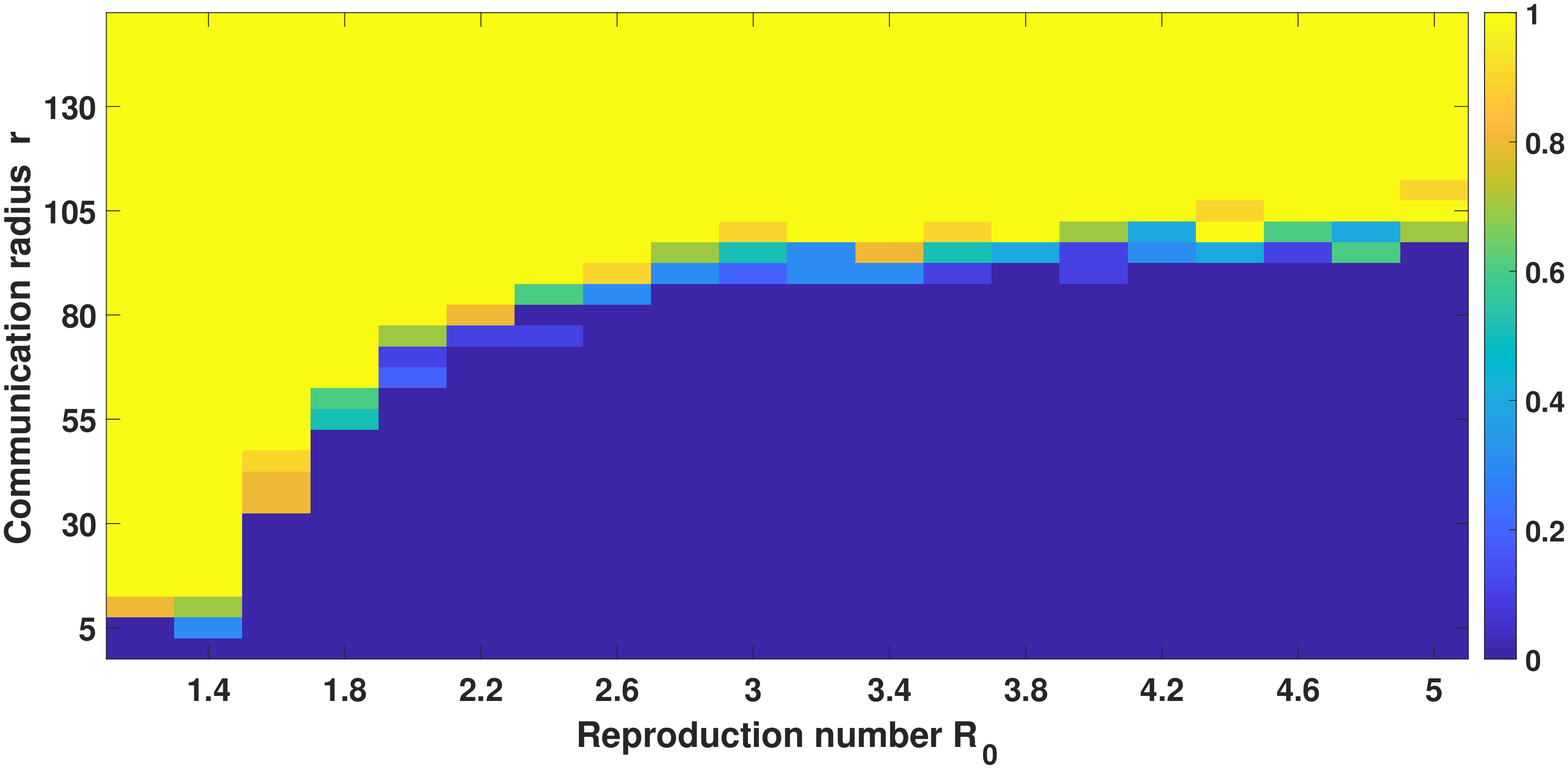}
      \vspace*{-2cm}}
  \caption{Success rates of resilient consensus
           versus the  $R_0$ of SIR model
           and $r$ in the regular agents}
 \label{fig.ggg3}
\end{figure*}

\textcolor{black}{We choose another local static policy for comparison. The policy maker chooses $b_0=b^*$ while the regular agents in this case set their pruning numbers individually according to the numbers of their neighbors as $F_{i0} = \lceil (1-b_0)\frac{d_i}{2} \rceil$. Compared to the local policy discussed above, this local version may be less resilient since the pruning number is smaller in general when the number of infected neighbors is large.
The heatmaps for homogeneous and gathered environments are displayed in Fig.~\ref{fig.ggg3}. From Fig.~\ref{fig.new7}, we observe that in a homogeneous environment, the policy works well in a wide range of $R_0$ and communication radius ($r\ge 80$).
In sparse networks, the part that has limited neighbors must have fewer infected nodes. The regular agents in such areas may remove less neighbors since $F_{i0} = \lceil (1-b_0)\frac{d_i}{2} \rceil$ and $d_i$ is small.
However, as shown in Fig.~\ref{fig.new8}, in the gathered infection environment, the proposed local static policy performs much worse. Since in the gathered areas, regular agents use the pruning numbers smaller than what are necessary, the agents fail to reach resilient consensus. The results for the local dynamic policy by choosing \eqref{eqn:bk} and $F_{i}(k) = \lceil (1-b_s(k))\frac{d_i}{2} \rceil$, which satisfies \eqref{eqn:Fi2}, but does not satisfy \eqref{eqn:Fi2_dyn}, have similar profiles as the local static policy and the related arguments also hold here.}


\vspace{-1 ex}

\subsection{Time Response for the Proposed Two Policies}
\begin{figure*}[htb]
  \centering
  \subfigure[Static policy with $R_0=5, r=70$]{%
      \label{fig.addnew1}
      \includegraphics[width=.48\linewidth]{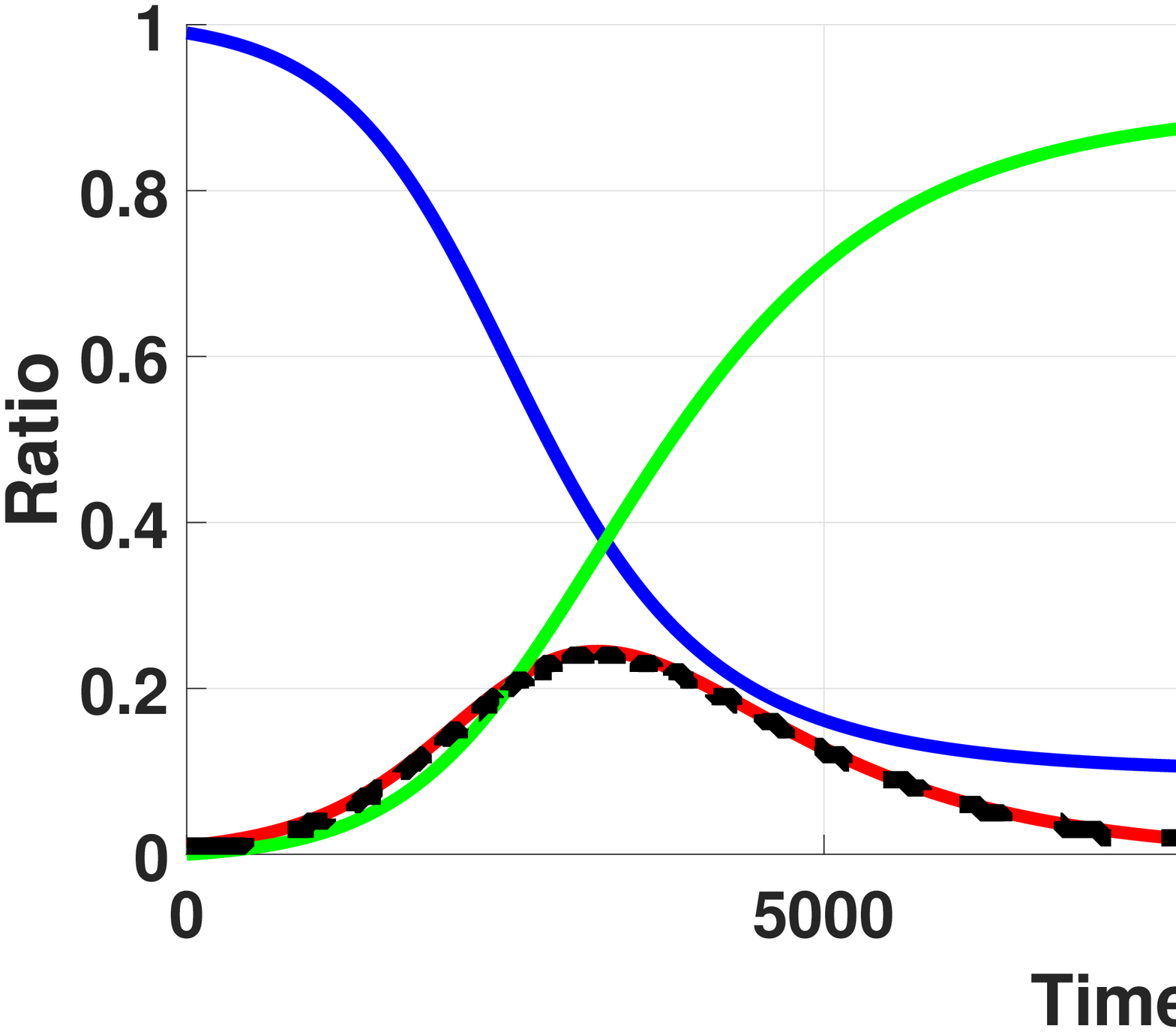}
      \vspace*{-2cm}}
  \subfigure[Dynamic policy with $R_0=5, r=70$]{%
      \label{fig.addnew2}
      \includegraphics[width=.48\linewidth]{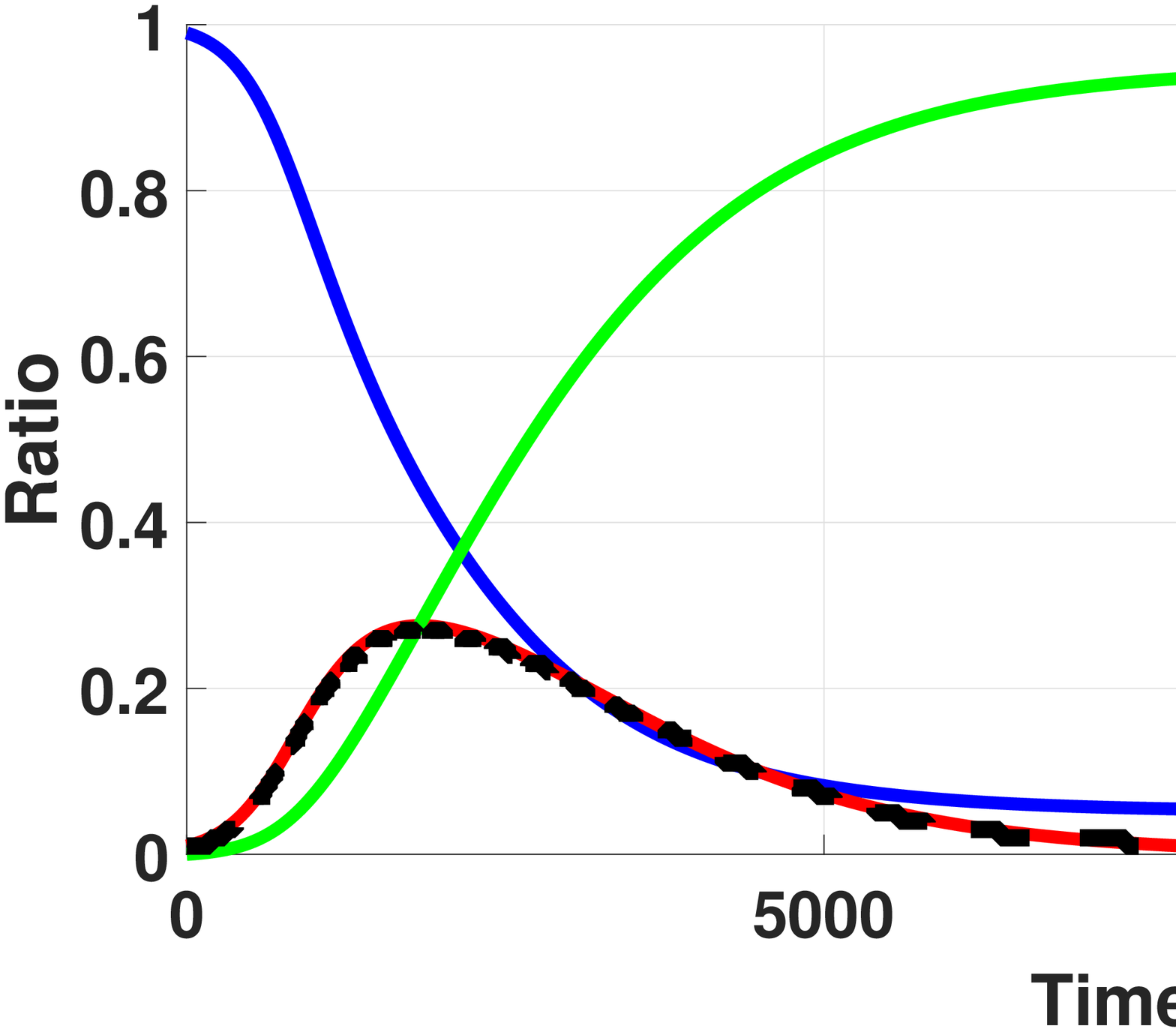}
      \vspace*{-2cm}}
  \subfigure[Static policy with $R_0=5, r=100$]{%
      \label{fig.addnew3}
      \includegraphics[width=.48\linewidth]{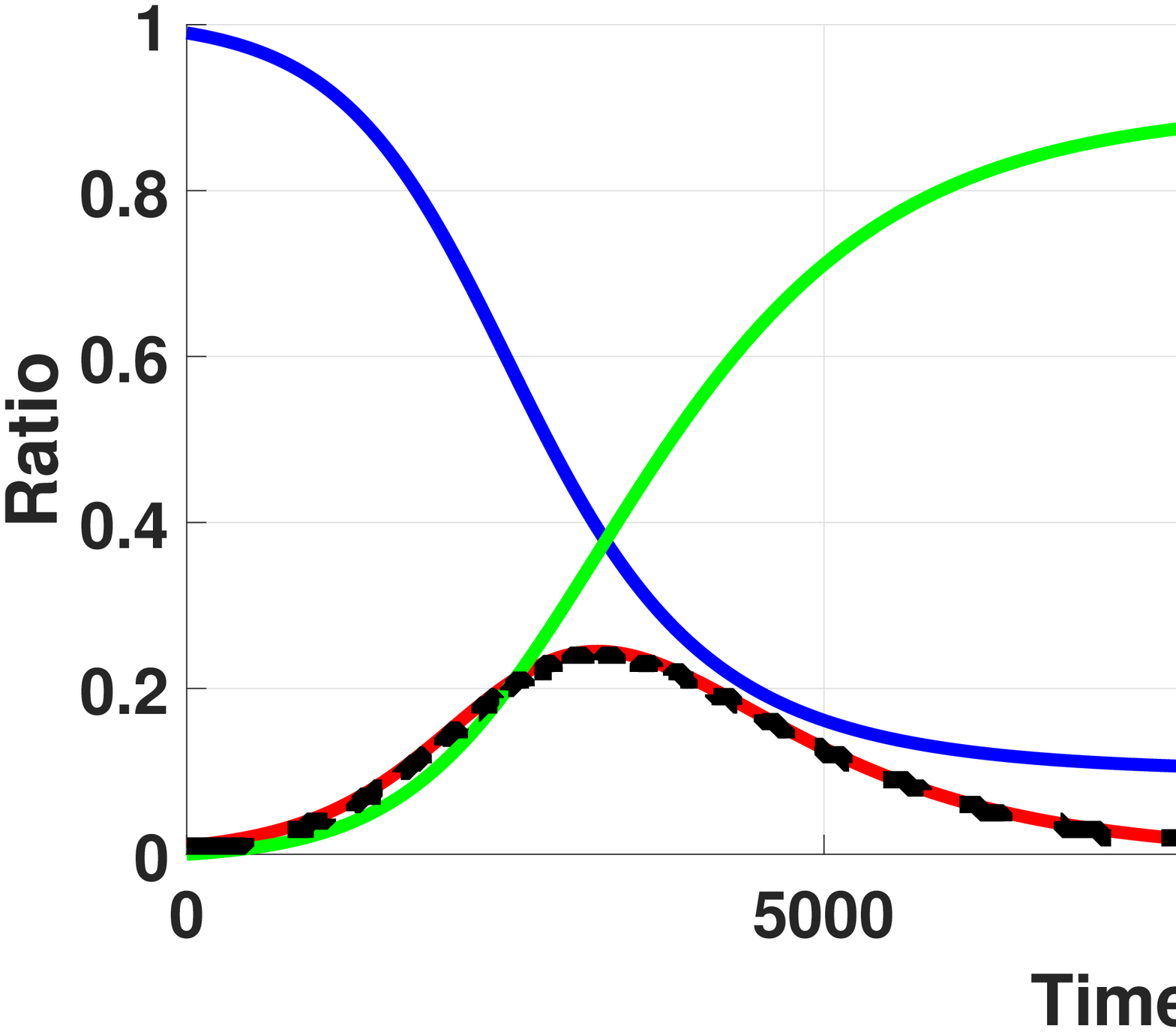}
      \vspace*{-2cm}}
  \subfigure[Dynamic policy with $R_0=5, r=100$]{%
      \label{fig.addnew4}
      \includegraphics[width=.48\linewidth]{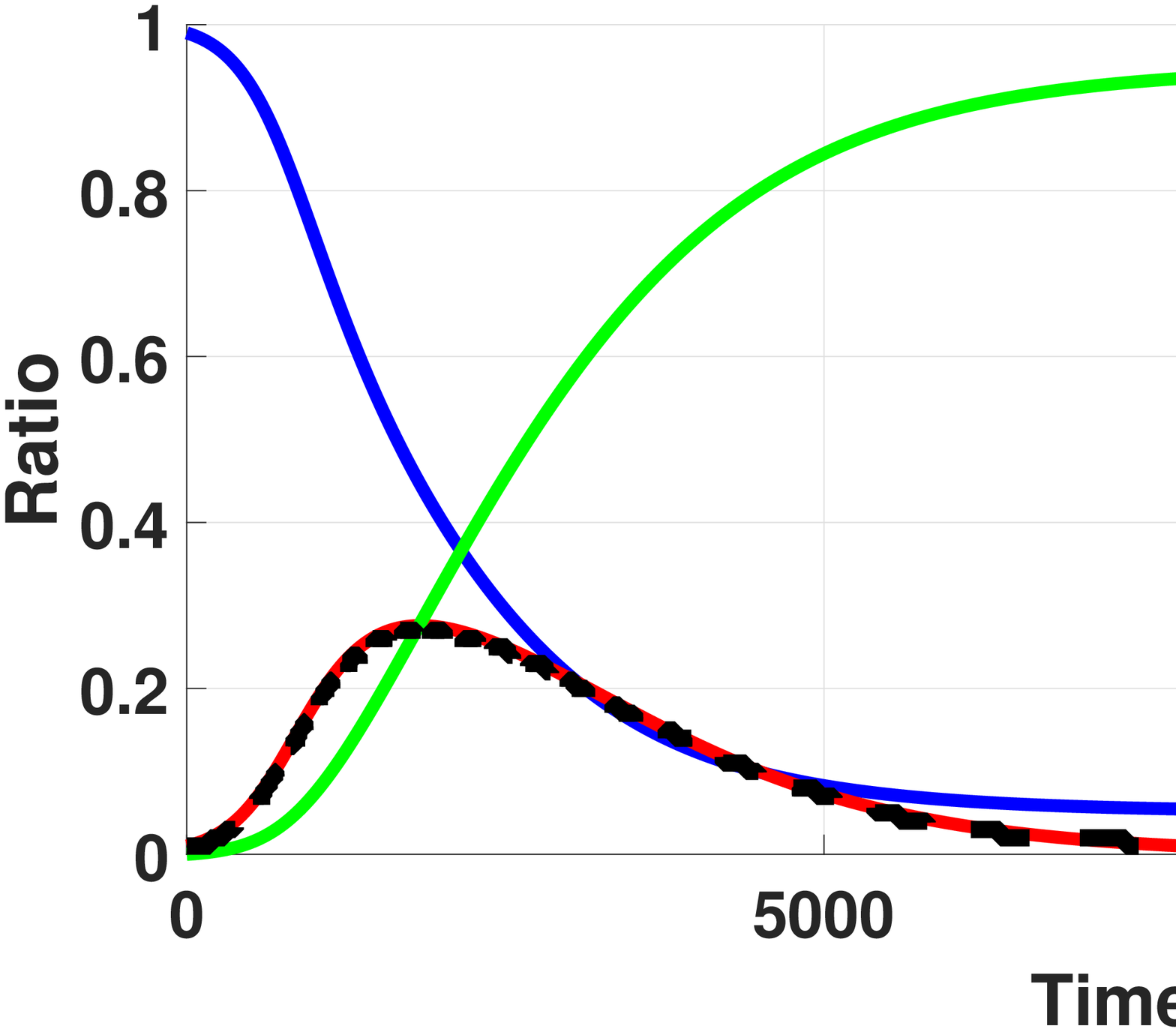}
      \vspace*{-2cm}}
  \subfigure[Local static policy with $R_0=5, r=70$]{%
      \label{fig.addnew5}
      \includegraphics[width=.48\linewidth]{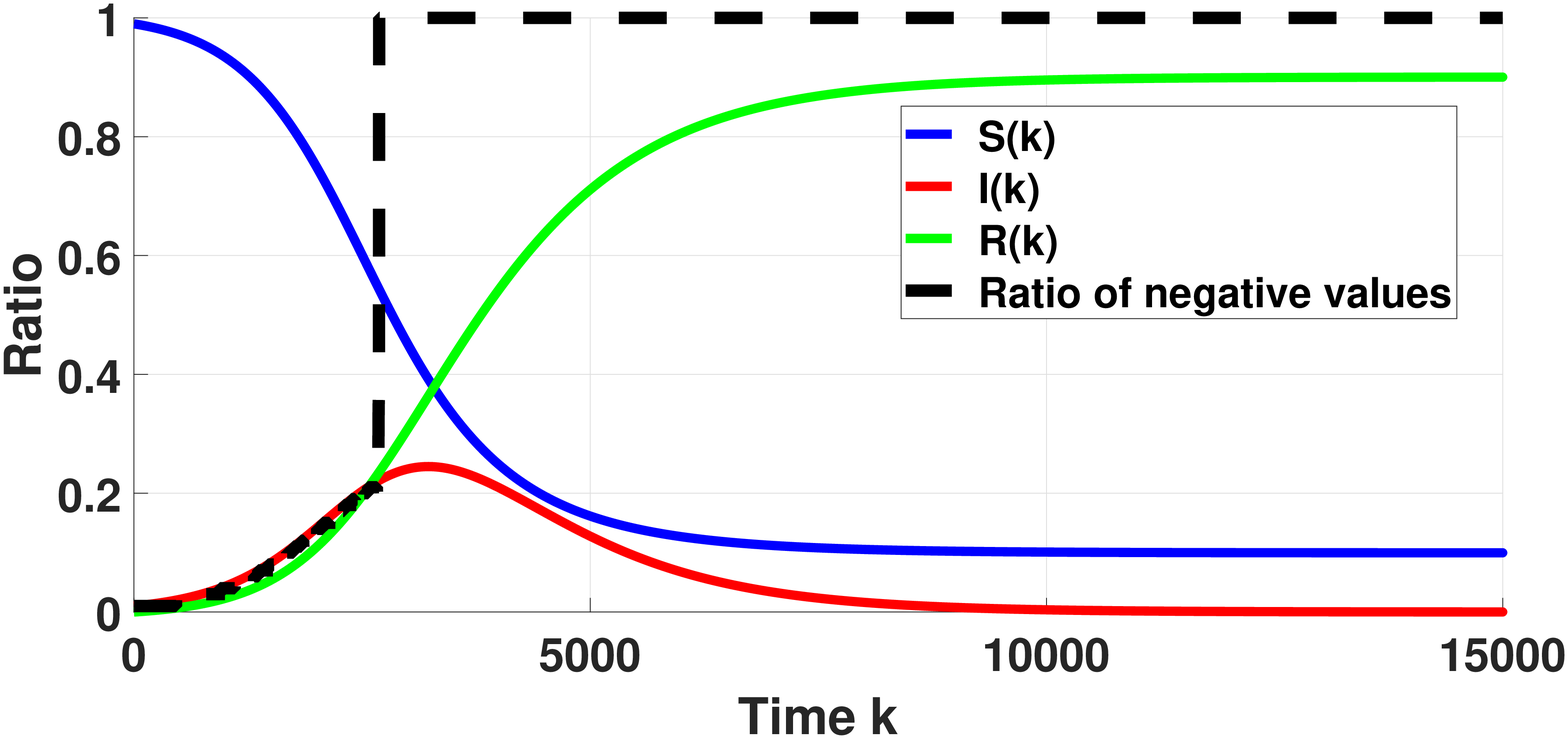}
      \vspace*{-2cm}}
  \subfigure[Local dynamic policy with $R_0=5, r=70$]{%
      \label{fig.addnew6}
      \includegraphics[width=.48\linewidth]{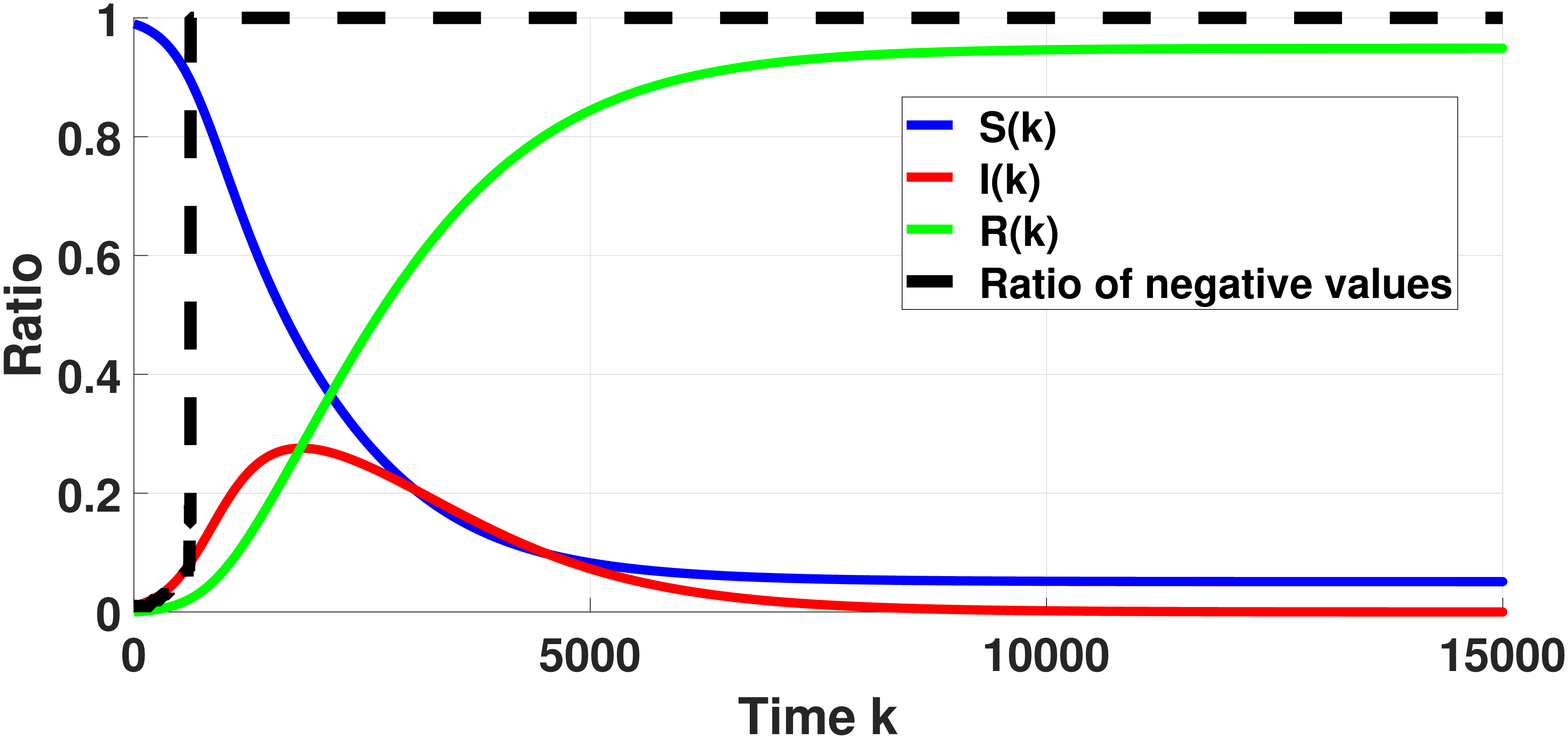}
      \vspace*{-2cm}}
  \subfigure[Local static policy with $R_0=5, r=100$]{%
      \label{fig.addnew7}
      \includegraphics[width=.48\linewidth]{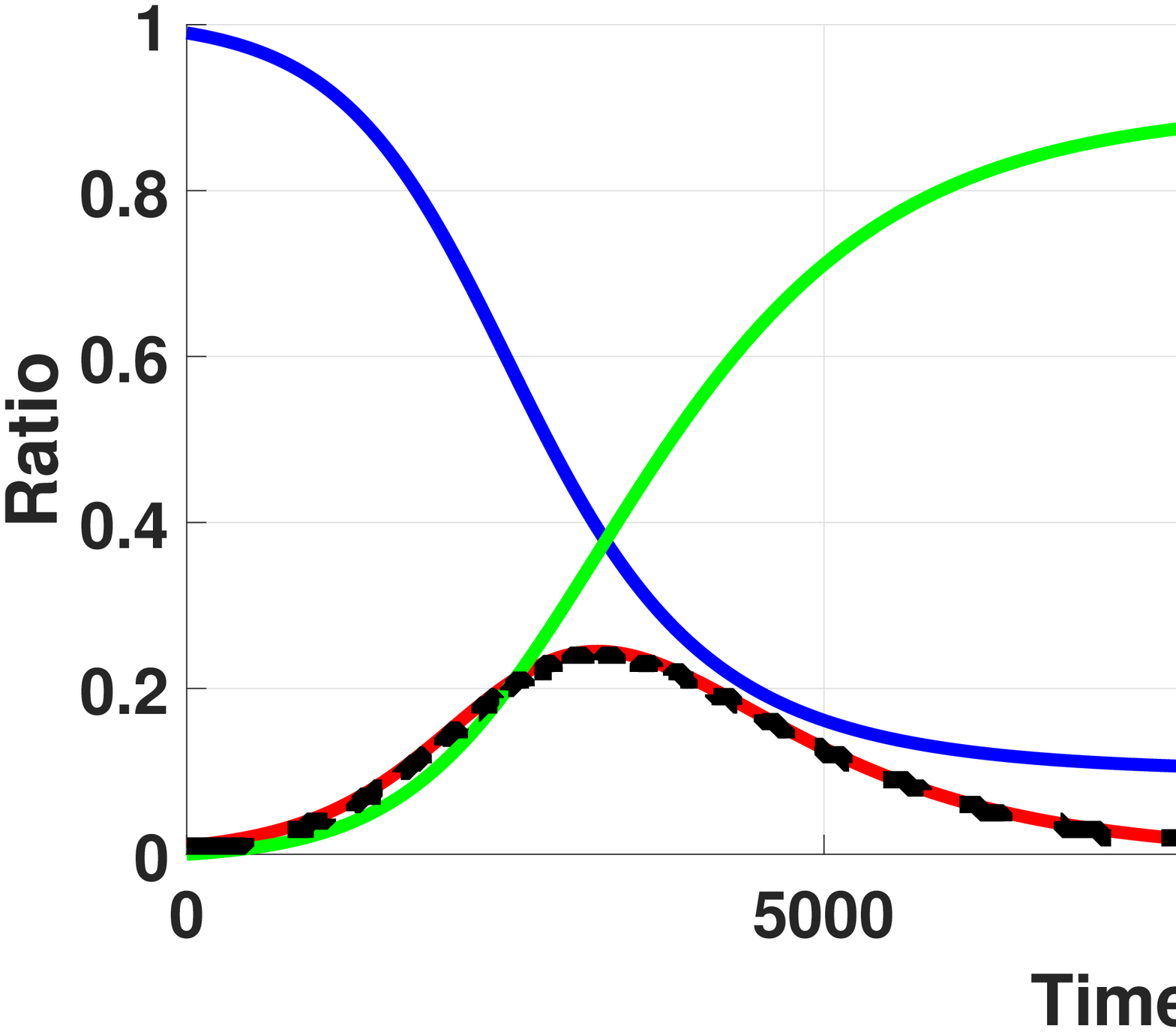}
      \vspace*{-2cm}}
  \subfigure[Local dynamic policy with $R_0=5, r=100$]{%
      \label{fig.addnew8}
      \includegraphics[width=.48\linewidth]{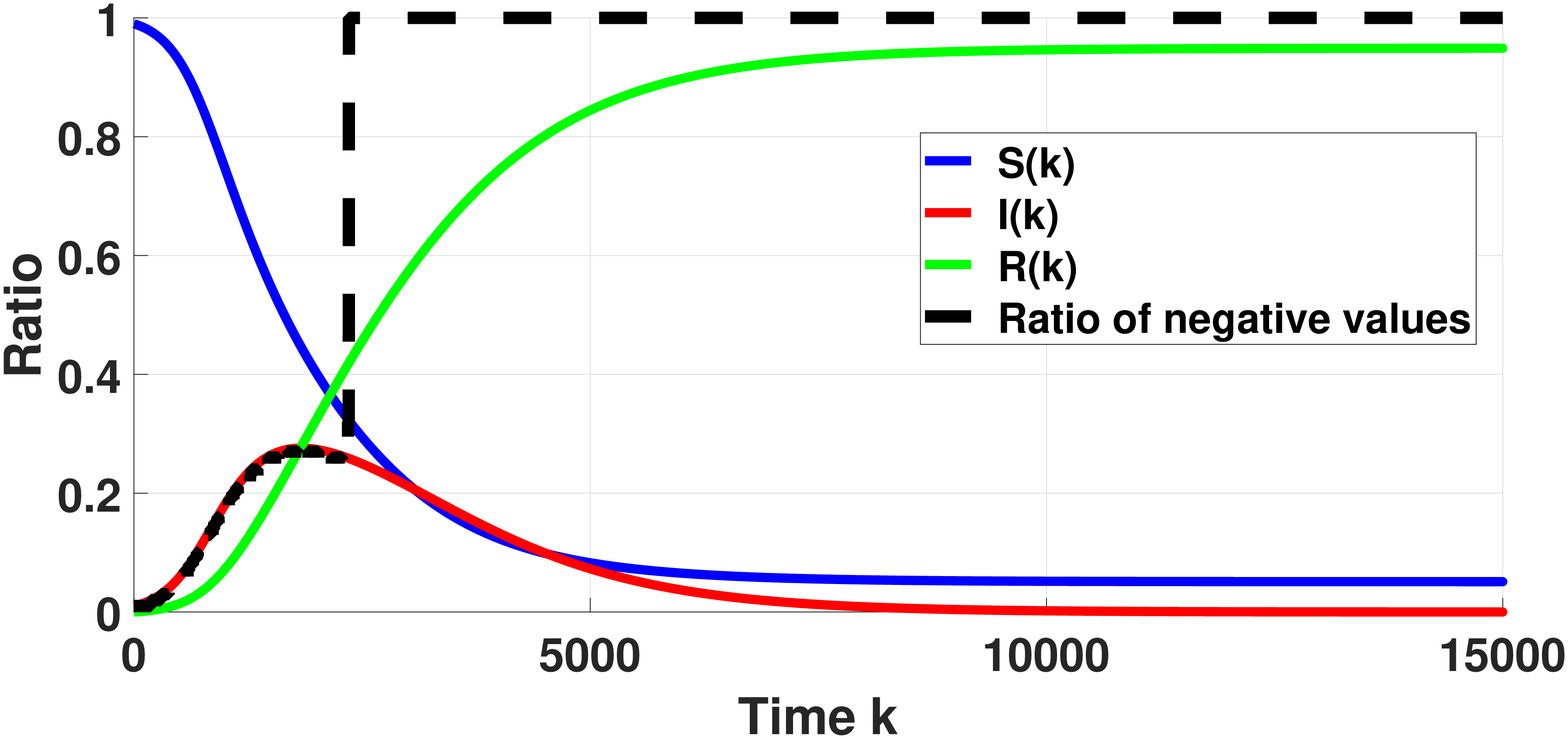}
      \vspace*{-2cm}}
  \caption{Time responses for different policies with $S(0)=0.99, I(0)=0.01, R(0)=0$}
 \label{fig.ggg4}
\end{figure*}

In this part of simulations, we check the time responses for the proposed static and dynamic policies. For this part, a small random network with 100 nodes is chosen in the area of $100 \times 100$. \textcolor{black}{Two local policy makers are placed in the two subgroups, where each subgroup contains 50 nodes.} The communication radius is 100.
The initial SIR ratio is $S(0)=0.99, I(0)=0.01$, and $R(0)=0$. The sampling period is $\Delta T = 0.01$. Susceptible and recovered agents randomly take initial values from $[0,1]$, which is considered as the safety interval. Once an agent is infected, its value is changed to $-1$ by the adversaries. The infected agents are randomly chosen, and hence the homogenous condition is not guaranteed.
The time responses for the two cases with $R_0 = 5, r=70$ and $R_0 = 5, r=100$ are shown in Fig.\ref{fig.ggg4}. We place the curves of $S(k),I(k), R(k)$ in the plots so that the real time infectious situation is clear. Here, we also plot the ratio of the agents taking negative numbers in the system. Under normal situations, this ratio should match that of the infected agents since the recovered agents will attain states within the safe interval shortly after their recovery. However, under severe conditions when the recovered agents have too many infected neighbors, this ratio may grow over time. We will see that a phase transition where this ratio becomes 1 can happen.

We first look at the proposed static and dynamic policy, shown in Figs.~\ref{fig.addnew1}, \ref{fig.addnew2}, \ref{fig.addnew3}, and \ref{fig.addnew4}. We can see that both policies guarantee resilient consensus and the infected values are almost all coming from the infected agents.
To compare the static and dynamic policies with the same $R_0$, the result shows that the dynamic policy has an earlier infectious peak. At the beginning, when $I(k)$ is low, the dynamic policy usually has fewer pruning numbers so that the peak may appear earlier. However, the $I_{\max}$ in the two policies is almost the same, which is also indicated in Fig.~\ref{fig.add1}.

Then, we look at the results for local static and dynamic policy, which are shown in Figs.~\ref{fig.addnew5}, \ref{fig.addnew6}, \ref{fig.addnew7} and \ref{fig.addnew8}. Within the sparse network ($r=70$), both local static and dynamic policies fail to reach resilient consensus. The regular agents fail to keep healthy states when the infected agents reach a certain level. Within the denser network ($r=100$), the local static policy reaches resilient consensus, but the local dynamic policy fails.
These results indicate that for both local static and dynamic policies, the homogenous condition is important for reaching resilient consensus. Moreover, compared with the local static policy, the local dynamic policy is more fragile since they remove less neighbor values. Once there is any agent who has more than $I(k)d_i/2$ infected neighbors, unsafe values dominate the agents, indicating that consensus is reached but it is not resilient.

\begin{figure*}[t]
  \centering
  \subfigure[Static policy with $R_0=5, r=100$]{%
      \label{fig.addnew9}
      \includegraphics[width=.48\linewidth]{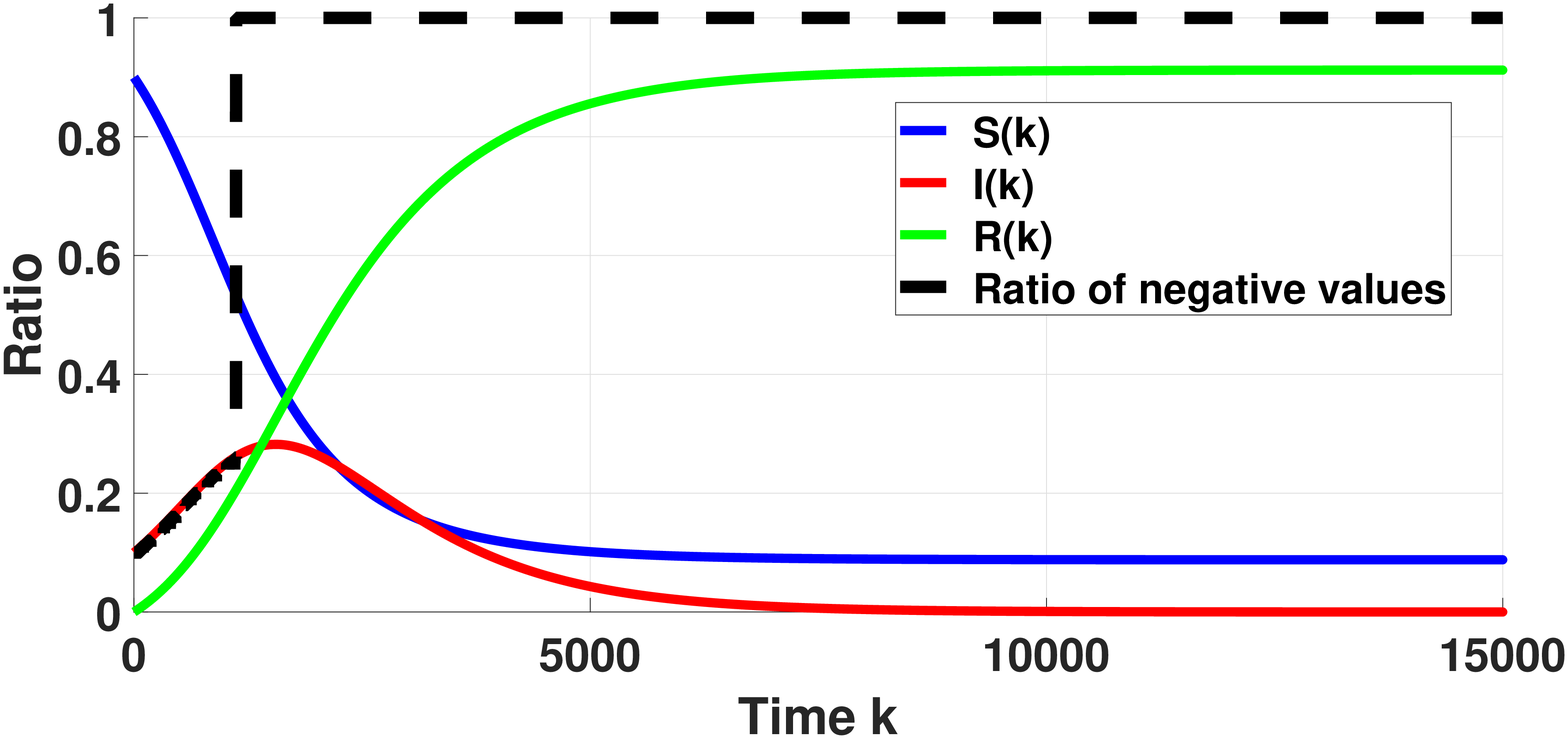}
      \vspace*{-2cm}}
  \subfigure[Dynamic policy with $R_0=5, r=100$]{%
      \label{fig.addnew10}
      \includegraphics[width=.48\linewidth]{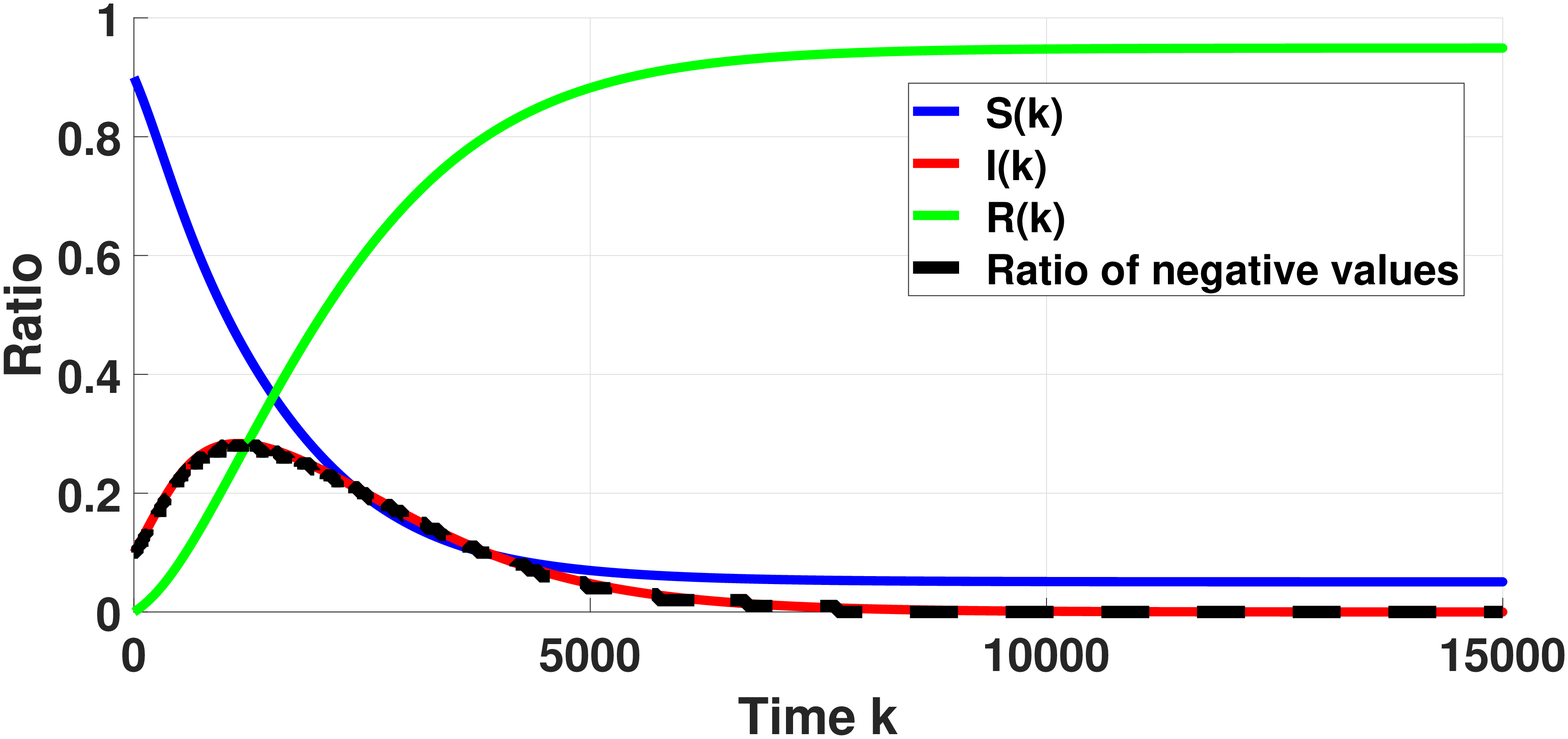}
      \vspace*{-2cm}}
  \subfigure[Local static policy with $R_0=5, r=100$]{%
      \label{fig.addnew11}
      \includegraphics[width=.48\linewidth]{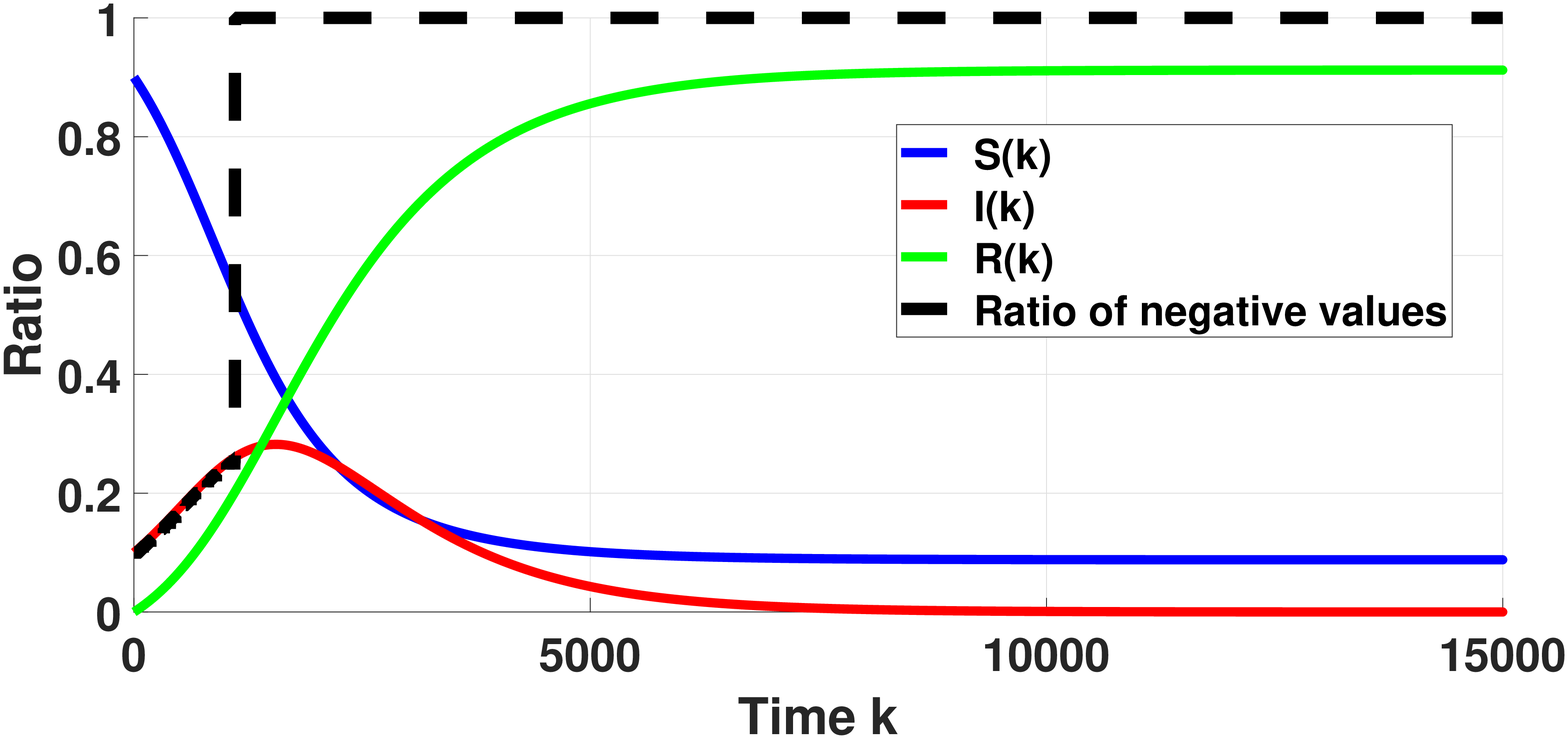}
      \vspace*{-2cm}}
  \subfigure[Local dynamic policy with $R_0=5, r=100$]{%
      \label{fig.addnew12}
      \includegraphics[width=.48\linewidth]{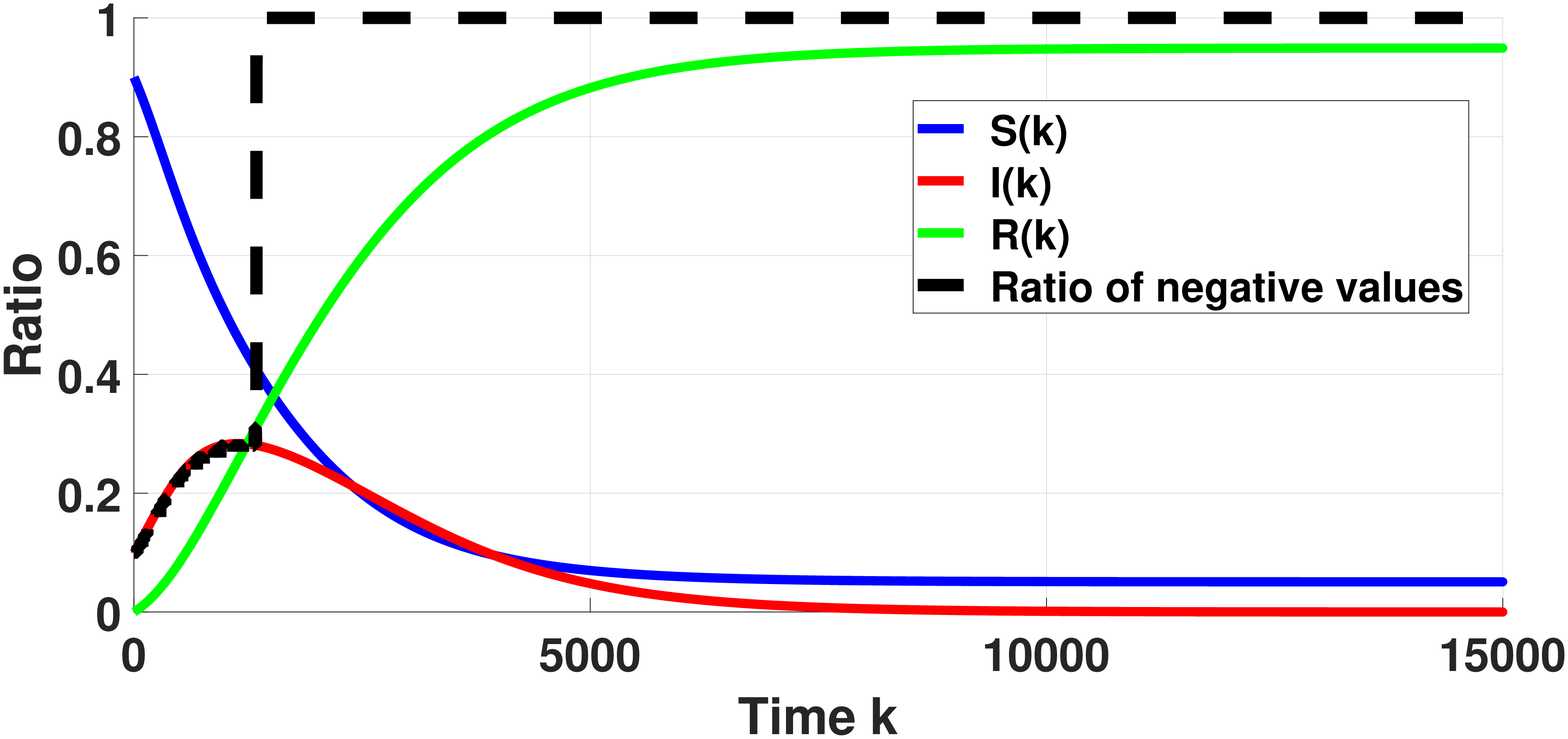}
      \vspace*{-2cm}}
  \caption{Time responses for different policies with $S(0)=0.9, I(0)=0.1, R(0)=0$}
 \label{fig.ggg5}
\end{figure*}

\vspace{-1 ex}

Next, we would like to show how the initial SIR ratios affect the performance of proposed two policies. We slightly changed the SIR ratios to $S(0)=0.9, I(0)=0.1, R(0)=0$ and generated the time responses for all policies. The results are shown in Fig.~\ref{fig.ggg5}. From the results in Figs.~\ref{fig.addnew9} and \ref{fig.addnew11}, we can see both the static policy and the local static policy fail to reach resilient consensus even within the dense network ($r=100$). The reason for the failure under the static policy comes from the approximate calculation for $I_{\max}$ in \eqref{eq-add01e}, where $S(0)\approx 1 , I(0) \approx 0$ are assumed. Here, we found that the initial ratios used here $S(0)=0.9, I(0)=0.1, R(0)=0$ do not satisfy the assumption so that the calculated $b = b^*$ does not match the real $I_{\max}$. When $I(k)$ increases over a bound (in this simulation it is $I(k) \ge 0.2$), there are more infected agents than the MSR algorithm is prepared for, preventing resilient consensus. On the other hand, for the dynamic policy, we do not have such an issue and the ratio of negative values reduced to zero together with the infectious ratio $I(k)$ as shown in Fig.~\ref{fig.addnew10}. We also found that if $S(0)\ge 0.97, I(0)\le 0.03, R(0)=0$, then the static policy can guarantee resilient consensus with $R_0=5$.



\vspace{-1 ex}

\section{Conclusion}
\label{Section 7}

In this paper, we have considered resilient consensus problems
in the presence of misbehaving agents, whose number changes
according to the level of pandemic.
Resilient protocols have been proposed to mitigate their influence
on regular agents.
Analyses have been made for both
static and dynamic policies for the transmission reduction parameter.
Moreover, we characterized the relation between
graph conditions and the pandemic reproduction number.
Numerical simulations further studied the theoretical and practical bound, homogeneous conditions and time responses
of the proposed protocols in random graphs.
Future directions include extending this study to other distributed epidemic models {and to multi-rate sampling settings,
and developing other measures for
reduction of contacts and social distancing
among agents. We would also like to extend our work to include more sophisticated models for the epidemics such as those incorporating network structures among subcommunities \textcolor{black}{\cite{Pare2017,Trajanovski2015}}. }

\vspace{-5ex}

\begin{IEEEbiography}[%
]{Yuan~Wang}(M'19)
Yuan Wang received the M.Sc.\ degree in engineering from
Huazhong University of Science and Technology, Wuhan,
China in 2016, and the Ph.D.\ degree in Artificial
Intelligence from Tokyo Institute of Technology, Yokohama,
Japan in 2019. He is currently a researcher in the Division of Decision and Control Systems,
KTH Royal Institute of Technology,
Stockholm, Sweden.

His main research interests are cyber-physical
systems, event-based coordination, security in multi-agent systems,
and model predictive control methods.
\end{IEEEbiography}

\vspace{-5ex}

\begin{IEEEbiography}[%
]{Hideaki Ishii} (M'02-SM'12-F'21) received the M.Eng.\ degree in
applied systems science from Kyoto University, Kyoto, Japan,
in 1998, and the Ph.D.\ degree in electrical and computer
engineering from the University of Toronto, Toronto, ON,
Canada, in 2002.
He was a Postdoctoral Research Associate with
the Coordinated Science Laboratory at the University
of Illinois at Urbana-Champaign, Urbana, IL, USA,
from 2001 to 2004, and a Research Associate with the
Department of Information Physics and Computing,
The University of Tokyo, Tokyo, Japan, from 2004 to 2007.
Currently, he is an Associate Professor in the Department
of Computer Science, Tokyo Institute of Technology, Yokohama, Japan.
His research interests are in networked control systems,
multiagent systems, cyber security of power systems,
and distributed and probabilistic algorithms.

Dr.~Ishii has served as an Associate Editor for
the \emph{IEEE Control Systems Letters} and
the \emph{Mathematics of Control, Signals, and Systems}
and previously for \emph{Automatica}, the \emph{IEEE Transactions
on Automatic Control}, and the \emph{IEEE Transactions on Control
of Network Systems}.
He is the Chair of the IFAC Coordinating Committee on Systems and Signals
since 2017.
He received the IEEE Control Systems Magazine Outstanding
Paper Award in 2015.
\end{IEEEbiography}

\vspace{-5ex}

\begin{IEEEbiography}[%
]{Fran\c{c}ois~Bonnet}
Fran\c{c}ois Bonnet is a Specially Appointed
Associated Professor at Tokyo Tech since 2018.

He obtained his M.S.\ from the ENS Cachan at Rennes,
France in 2006 and his Ph.D.\ from the University of
Rennes~1 in 2010. He worked at JAIST as a JSPS postdoctoral
fellow until 2012 and then as an Assistant Professor until 2017.
Then he spent one year as a Specially Appointed Assistant
Professor at Osaka University.

His research interests include theoretical distributed
computing, discrete algorithms, and (combinatorial) game
theory.
\end{IEEEbiography}

\vspace{-5ex}
\begin{IEEEbiography}[
]{Xavier~D\'{e}fago}(M'93)
Xavier D\'{e}fago is a full professor at Tokyo
Institute of Technology since 2016.

He obtained master (1995) and PhD (2000) in computer science from the Swiss Federal Institute of Technology in Lausanne (EPFL) in Switzerland. Before his current position at Tokyo Tech, he was a faculty member at the Japan Advanced Institute of Science and Technology (JAIST).
Meanwhile, he has also been a PRESTO researcher for the Japan Science and Technology Agency (JST), and an invited researcher for CNRS (France) at Sorbonne University and at INRIA Sophia Antipolis.

He is a member of the IFIP working group 10.4 on dependable computing and fault-tolerance. He served as program chair of IEEE SRDS in 2014 and IEEE ICDCS in 2012 and as general chair of SSS 2018.

Xavier has been working on various aspects of dependable computing such as distributed agreement, state machine replication, failure detection, and fault-tolerant group communication in general.
His interest include also robotics, embedded systems, and programming languages.
\end{IEEEbiography}

\end{document}